\documentclass[prc,aps,eqsecnum,epsf,amssymb]{revtex4}
\usepackage{graphicx}
\usepackage{bm}
\def\a{\alpha}
 
\def\htm{{\hbar^2 \over m}}

\def\tri{{{}^3{\rm H}}}
\def\hel{{{}^3{\rm He}}}
\def\het{{{}^3{\rm He}}}
\def\heq{{{}^4{\rm He}}}

\def\jac{\xi}
\newcommand{\jacb}{{\mbox{\boldmath $\xi$}}}
\def\hypfi{\varphi}
\def\pp{\phantom{(31.358)}}
\def\ppp{\phantom{4(31.358)}}
\newcommand{\bmsi}{{\mbox{\boldmath $\sigma$}}}

\newcommand{\bmta}{{\mbox{\boldmath $\tau$}}}
\newcommand{\bfr}{{\bm r}}
\begin{document}

\title{ Calculation of the $\alpha$--Particle  Ground State within the
Hyperspherical Harmonic Basis}
\author{M. Viviani$^1$, A. Kievsky$^1$ and S. Rosati$^{1,2}$}
\affiliation{ $^1$Istituto Nazionale di Fisica Nucleare, Via Buonarroti 2,
56100 Pisa, Italy }
\affiliation{ $^2$Dipartimento di Fisica, Universita' di Pisa, Via
Buonarroti 2, 56100 Pisa, Italy }
\date{\today}

\begin{abstract}
The problem of calculating the four--nucleon bound state
properties for the case of realistic two- and three-body 
nuclear potentials is studied
using the hyperspherical harmonic (HH) approach. 
A careful analysis of the convergence of different classes of HH
functions has been performed. A restricted basis is chosen to allow for 
accurate estimates of the binding energy and other properties of the
$\heq$ ground state. Results for various modern two-nucleon and 
two- plus three-nucleon interactions are presented. The $\heq$ 
asymptotic normalization constants for separation in 2+2 and 1+3
clusters are also computed.
\end{abstract}

\maketitle

\section{Introduction}
\label{sec:intro} 
Rapid progress has been made during the last few years in the
quantitative study of the $A=4$ nuclear systems. Ever increasing
computer power, development of novel numerical methods, and
significant refinements of well-established techniques have allowed
the solution of the four-nucleon bound state problem with a control of
the numerical error at the level of $10$-$20$ keV (the experimental 
$\a$-particle binding energy (BE) being $28.30$ MeV), 
at least for Hamiltonians including only nucleon-nucleon (NN) interaction
models~\cite{Kea01}. In the latter work, the BE  and other properties of the
$\a$-particle were studied with the AV8$'$~\cite{AV18+} NN
interaction, and the different techniques produced results in very
close agreement with each other (at the level of less than 1\%).

In Refs.~\cite{NKG00,Wea00}, realistic potential models have been used to
describe the $\alpha$-particle bound state. Those potential models
consist of the sum of a modern NN interaction plus a three-nucleon (3N)
interaction. A modern NN interaction has the property of
describing the NN database with a $\chi^2$ per datum close to one.
Examples are the Argonne V18 (AV18) potential~\cite{AV18}, the 
Nijmegen potentials~\cite{Nijm}, the
CD-Bonn potential~\cite{Bonn,BonnCD} and the recently proposed potential,
non-local in $r$-space, developed by Doleschall and coll.~\cite{DB00,Dea03}. 
This last potential has the remarkable property of reproducing 
simultaneously the NN bound and scattering data and the 3N binding 
energies. As is well known, the other models (AV18, Nijmegen 
and CD-Bonn) underbind the 3N system. Usually a 3N interaction is 
included in the Hamiltonian when these potentials are considered. 
The strength of the 3N interaction is properly tuned to reproduce 
the $\tri$  binding energy and this strength depends on the chosen 
NN potential.
Examples of 3N interactions are the Urbana IX (UIX)~\cite{AV18+}, the
Tucson-Melbourne (TM)~\cite{TM} and Brazil~\cite{BR} potentials.
From Ref.~\cite{NKG00} we observe that all the NN+3N potential models which
reproduce the deuteron and the 3N binding energies slightly overbind
the $\a$--particle. We further observe that the results obtained
using different techniques~\cite{NKG00,Wea00} for the AV18+UIX 
potential model though close to each other are not in complete agreement. 
Clearly a clarification of these points would be welcome.
Moreover, in recent years 
there has been a rapid progress in developing new models of
the nuclear interaction based on the application of the chiral perturbation
theory~\cite{W90,vanK94,EGM00}. From these studies one can hope 
to have a better understanding of the form of the NN and 3N interactions 
(the four-nucleon force is expected to be very small). All these 
potential models have to be studied in detail in the $A=3$ and $A=4$ 
systems. It is therefore very important to have powerful techniques 
for solving four-nucleon problems.

The methods devised to tackle the problem of the solution of 
the non-relativistic Schr\"odinger equation
\begin{equation}
  H\Psi(1,2,3,4) =E\Psi(1,2,3,4)\ ,
  \label{eq:es}
\end{equation}
where $H$ is the four-body nuclear Hamiltonian,  are very
different. In the Faddeev--Yakubovsky (FY)
approach~\cite{Y67,KG92,SSK92,NKG00},
Eq.~(\ref{eq:es}) is transformed to a set of coupled equations for the
FY amplitudes, which are then solved directly (in momentum or
coordinate space) after a partial wave expansion.
In the Green Function Monte Carlo (GFMC) method~\cite{C88,Wea00}
one computes $\exp(- \tau H)\Phi(1,2,3,4)$, where $\Phi(1,2,3,4)$ is a 
trial wave function (WF), using a stochastic procedure to obtain,
in the limit of large  $\tau$, the exact ground state WF $\Psi$.
These two techniques have also been applied to the case where the nuclear
Hamiltonians includes a 3N interaction.
The Stochastic Variational Method (SVM)~\cite{VS95,SV98}
and the Coupled Rearrangement Channel Gaussian-Basis method
(CRCG)~\cite{KKF89,KK90}
provide a variational solution of Eq.~(\ref{eq:es}) by expanding the
(radial part of the) WF in gaussians. The two techniques differ in the way
they determine the non-linear coefficients of
the expansion: 
in the SVM random choices are used to select the optimum set, whereas
in the CRCG technique the non linear coefficients are chosen in geometrical
progression in such a way that only a few of them have to be varied.
Very recently two other new techniques have been
proposed. In the no-core shell model (NCSM) method~\cite{NB99,NVB00}
the calculations are performed using a (translational-invariant)
harmonic-oscillator (HO) finite basis $P$ and introducing an effective
$P$-dependent Hamiltonian $H_{P}$  to replace $H$ in
Eq.~(\ref{eq:es}).
The operator $H_{P}$ is constructed so that the solution of
the equation $H_{P}\Psi(P)= E_{P} \Psi(P)$ provides  eigenvalues
which quickly converge to the exact ones as $P$ is enlarged.  The
effective interaction hyperspherical harmonic (EIHH) method~\cite{BLO00}
is based on a similar idea, but the finite basis $P$ is
constructed in terms of the hyperspherical harmonic (HH) functions.

In the present work we would like to address the problem of
calculating the $\alpha$-particle properties, using a nuclear
Hamiltonian containing modern two- and three-nucleon interactions,
by expanding the WF in terms of the HH functions.
Our intention is to obtain converged binding energies at the level
of $20$--$30$ keV. The motivation is twofold. First we would like
to reduce the theoretical error in the determination of the $\a$--particle
bound state properties. Second, the HH techniques
can also be extended to treating four nucleons scattering states,
as has been possible for the $A=3$ system~\cite{KRV94} using a similar
technique. This program is currently under way and a preliminary report 
has been already published~\cite{KRV04}. The richness of 
phenomena in the four nucleon scattering and reactions will be an ideal 
laboratory for studying and testing newer models of the 
nuclear interaction.

In an earlier work~\cite{KRV95}, the authors determined 
the solution of Eq.~(\ref{eq:es}) variationally 
by expanding the WF in a basis
of correlated Hyperspherical Harmonic (CHH) WF's. The space
part of such a basis consisted of products of correlation factors 
$F$ and HH functions. The correlation factors $F$ were chosen so as 
to take into account the strong correlations induced by
the NN potential, especially at short inter-particle distances.
The introduction of such factors substantially improved the 
convergence of the expansion. This made it possible to obtain reasonable 
estimates for the ground state energy of the $\alpha$ particle and some 
selected observables in $n-\tri$ and 
$p-\hel$ elastic scattering using a rather 
limited basis set~\cite{KRV95,KRV98,KRV01}. 
However, due to the complexity of $F$, 
the spatial integrations were performed by using quasi-random number
techniques. The precision of the required matrix elements was 
therefore limited, and the inclusion of a greater number of states 
was problematic.

When the four--nucleon WF is expanded in terms of the uncorrelated HH
basis (i.e. setting $F=1$) most of the integrations can be performed
analytically, and the remaining low-dimensional integrals can be
evaluated by means of efficient quadrature methods.
However, due to the particular structure of the NN potential, which is
state dependent and strongly repulsive at short distances, a very large
number of basis elements are required. For that reason, the application of 
the HH technique to studying the $A=4$ nuclear system has encountered serious 
convergence problems. Few four--body HH calculations have been attempted so 
far for realistic interactions~\cite{DPE73,FE81,B86}. 
Even for central or super-soft-core potentials the problem of 
the slow convergence of the HH expansion has not been completely 
overcome~\cite{DPE73,B86,B81}. 
The reason for these difficulties is related to the slow convergence of
the basis with respect to the grand angular quantum number $K$ and
to the large number of HH states with a given $K$. For example, for
an accurate description of the $\alpha$-particle ground state,
antisymmetric spin-isospin-HH states up to $K=60$ 
have to be included. However, the number of such states already
for $K=20$ is greater than 1,000 and it increases very rapidly
with $K$. It is therefore clear that a
{\it brute force} application of the method is not possible even
with sophisticated computational facilities.

The approach analyzed in the past was to select a suitable
subset of states~\cite{E72,D77,F83}. In those papers it resulted 
quite clearly that the quantum
number $K$ is not the unique parameter important for studying the
convergence of the basis. Let us recall that a 4-body HH function is
specified by
three orbital angular momentum quantum numbers $\ell_1$, $\ell_2$,
$\ell_3$ and two additional quantum numbers $n_2$, $n_3$ (which
are non--negative integers) related to the radial excitation of
the system. The grand angular quantum number is defined to be
$K=\ell_1+\ell_2+\ell_3+2(n_2+n_3)$. Note that ${\cal
L}=\ell_1+\ell_2+\ell_3$ and $K$ are even (odd) numbers for positive
(negative) parity states. In Ref.~\cite{E72}, the
basis was restricted  to including HH states with a few choices of
$\ell_1$, $\ell_2$, $\ell_3$ values, and large values of
$n_2$ and $n_3$.  The calculations performed~\cite{DPE73,FE81} were however
limited by the computer power available at that time. In this paper,
it is shown that HH states having
${\cal L}\equiv\ell_1+\ell_2+\ell_3\le 6$ are sufficient in order
to obtain a four digit convergence. However, the number of HH
states with ${\cal L}{\le 6}$ is still huge and additional criteria
for selecting a reduced basis have to be specified.

It is possible to organize the HH states in
terms of the number of particles correlated. For example, there is
a class of elements which depends only on the coordinate of two
particles, the so called Potential Basis (PB)~\cite{F83}. Such a basis
therefore takes the two--body correlations into account.
However, even in the case of simple model interactions, the
BE's $B$ obtained by restricting the expansion basis to
the PB were found to be rather far from the exact values. For
example, for the Malfliet-Tjon V (MT-V) central potential~\cite{MT69},
$B$ calculated with the PB is approximately 1 MeV smaller than the
exact value. For a realistic potential the situation is noticeably worse. 
However, it is clear
that the procedure of classifying the HH states in terms of the number
of correlated particles can be useful for distinguishing the 
importance of the various expansion terms.

In the present paper the application  of the HH expansion basis
is developed by taking advantage of both 
strategies discussed above. Namely, HH states of low values of $\ell_1$,
$\ell_2$, $\ell_3$ are included first. Among them,
those correlating only a particle pair are included first, then those
correlating three particles are added and so on. In practice, the HH 
states are first divided into {\it classes} depending on
the value of ${\cal L}$ and $n_2$, $n_3$. 
Let us denote with ${\cal M}_i(K_i)$ the number of states 
belonging to a class $i$ with $K\le K_i$. Such a number of states
rapidly increases with $K_i$, in general ${\cal M}_i(K_i)\approx  M_i
(K_i)^r$ for large $K_i$, where the numbers $M_i$ are a set of constants 
and $r=1$ or $2$.
The first and most important classes should contain only a 
small subset of HH states (``small'' classes), namely their
$M_i$ should be small, let us say $M_i\approx 1$. It is then relatively
easy to include HH states of large $K$ belonging to these classes.
The classes containing successively larger numbers of states $(M_i\gg 1)$
should be chosen possibly so as to contribute lesser and lesser to the 
expansion in such a way that the
corresponding expansion can be truncated to small values of $K$.
A careful analysis of the convergence properties of the various 
HH components has allowed for an optimal choice of the classes,
so that accurate calculations of the $\alpha$-particle properties
could be achieved.

An important aspect of a successful application of the HH method 
is related to the computation of the
coefficients for the transformation of a HH function corresponding
to a generic permutation of the 4 particles in terms of those
constructed for a given permutation. Various approaches have been
devised to deal with this
problem~\cite{K72,E79,NK94,DLWD95,E95,V98}. The usefulness of these
coefficients is twofold. First, it is easy to identify 
the linearly dependent states and to avoid their inclusion in the
expansion basis. The removal of these ``spurious" states, which
disappear after a proper antisymmetrization of the basis, is very
useful as the number of linear independent states is noticeably
smaller than the full degeneracy of the basis.
Secondly, the matrix elements of a local two-body (three-body) potential 
energy operator are easily reduced to one-dimensional (tri-dimensional)
integrations, which can also be performed beforehand and stored on
computer disks. The matrix elements of non-local operators can also be 
reduced to low-dimensional integrals.
The kinetic energy operator is easily obtained analytically.

This study is the continuation of the application of the HH
expansion to the three nucleon system performed in
Ref.~\cite{KMRV96}. 
Another possible extension of these studies are related  to the
application of the HH technique to heavier systems. In particular,
we can point out that also for $A>4$ the calculation of the
multidimensional integrals related to the matrix elements of a
local  NN (3N) interaction reduces to a one-dimensional 
(three-dimensional) integration.
The only difficulty in extending the method to heavier systems is the
choice of a suitable and optimized subset of HH functions. We hope
that the criteria used here to select an optimal subset of the basis
could also be applied for systems with $A>4$.

This paper is organized as follows. In the next section, a brief
description of the properties of the HH functions is reported. In section 3,
the choice of the basis is presented. The results obtained for the
BE and other properties of the $\alpha$-particle
are presented in Sect.4. Finally, the last
section is devoted to the conclusions and
the perspectives of the present approach.

\section{The HH expansion}
\label{sec:hh}

For four equal mass particles, a suitable choice of the Jacobi
vectors is
\begin{eqnarray}
          \jacb_{1p}& = & \sqrt{3\over2} (\bfr_m-
           { \bfr_i+\bfr_j +\bfr_k\over 3} )\ , \nonumber\\
         \jacb_{2p} & = & \sqrt{4\over3} (\bfr_k-
             { \bfr_i+\bfr_j \over 2} )\ ,  \label{eq:JcbV}\\
         \jacb_{3p} & =& \bfr_j-\bfr_i\ , \nonumber
\end{eqnarray}
where $p$ specifies a given permutation corresponding to the order $i$, $j$,
$k$ and $m$ of the particles. By definition, the permutation $p=1$ is chosen
to correspond  to the order $1$, $2$, $3$ and $4$.

For a given choice of the Jacobi vectors, the hyperspherical
coordinates are given by the hyperradius $\rho$, defined
by
\begin{equation}
   \rho=\sqrt{\jac_{1p}^2+\jac_{2p}^2+\jac_{3p}^2}
    \ ,\qquad ({\rm independent\    on\ }p)\ ,
    \label{eq:rho}
\end{equation}
and by a set of variables which in the Zernike
and Brinkman~\cite{zerni,F83} representation are the polar
angles $\widehat \jac_{ip}\equiv (\theta_{ip},\phi_{ip})$  of each Jacobi
vector, and the two additional ``hyperspherical'' angles
$\hypfi_{2p}$ and $\hypfi_{3p}$ defined as
\begin{equation}
    \cos\hypfi_{2p} = { \jac_{2p} \over \sqrt{\jac_{1p}^2+\jac_{2p}^2}}\ ,
    \qquad
    \cos\hypfi_{3p} = 
    { \jac_{3p} \over \sqrt{\jac_{1p}^2+\jac_{2p}^2+\jac_{3p}^2}}
     ={\jac_{3p}\over \rho}\ ,
     \label{eq:phi}
\end{equation}
where $\jac_{jp}$ is the magnitude of the Jacobi vector $\jacb_{jp}$. The set
of the variables $\widehat \jac_{1p},\widehat \jac_{2p},\widehat \jac_{3p},
\hypfi_{2p}, \hypfi_{3p}$ is denoted  hereafter as $\Omega_p$. To
simplify the notation for $p=1$, the subscript ``$1$" will be
sometime omitted. The expression of a generic HH function is
\begin{equation}
  {\cal Y}^{K,LM}_{\ell_1,\ell_2,\ell_3, L_2 ,n_2, n_3}(\Omega_p)  =
   \left [ \Bigl ( Y_{\ell_1}(\widehat \jac_{1p})
    Y_{\ell_2}(\widehat \jac_{2p}) \Bigr )_{L_2}  
    Y_{\ell_3}(\widehat \jac_{3p}) \right
    ]_{LM} {\cal P}^{\ell_1,\ell_2,\ell_3}_{n_2, n_3}
     (\hypfi_{2p},\hypfi_{3p}) \ ,\label{eq:hh4}
\end{equation}
where
\begin{eqnarray}
 {\cal P}^{\ell_1,\ell_2,\ell_3}_{n_2, n_3}(\hypfi_{2p},\hypfi_{3p}) &= &
 {\cal N}^{\ell_1,\ell_2,\ell_3}_{ n_2, n_3}
   \sin^{\ell_1 }\hypfi_{2p}    \cos^{\ell_2}\hypfi_{2p}
   \sin^{\ell_1+\ell_2+2n_2}\hypfi_{3p}
      \cos^{\ell_3}\hypfi_{3p} \times    \nonumber \\
   &&
      P^{\ell_1+{1\over 2}, \ell_2+{1\over 2}}_{n_2}(\cos2\hypfi_{2p})
      P^{\ell_1+\ell_2+2n_2+2, \ell_3+{1\over 2}}_{n_3}(\cos2\hypfi_{3p})\ ,
      \label{eq:hh4P}
\end{eqnarray}
and $P^{a,b}_n$ are Jacobi polynomials.
The coefficients ${\cal N}^{\ell_1,\ell_2,\ell_3}_{ n_2, n_3}$ are
normalization factors, given explicitly by
\begin{equation}
  {\cal N}^{\ell_1,\ell_2,\ell_3}_{ n_2, n_3} =
   \prod_{j=2}^3 \left [ { 2\nu_j \Gamma(\nu_j-n_j) n_j! \over
   \Gamma(\nu_j-n_j-\ell_j-1/2) \Gamma(n_j+\ell_j+3/2) } \right
   ]^{1\over 2}\ ,
   \label{eq:norma4}
\end{equation}
where $\nu_j=K_j+(3j-5)/2$ with $K_j$  defined to be
\begin{equation}
    K_2= \ell_1+\ell_2+2 n_2\ , \qquad
    K_3=K_2+\ell_3+2 n_3 \equiv K \ ,
     \label{eq:go}
\end{equation}
and $K$ is the grand angular quantum number.

The HH functions are eigenfunctions of the hyperangular part of
the kinetic energy operator $\Lambda^2$. In fact, for $A=4$
the latter operator can be written using the variables
$\{\rho,\Omega_p\}$ as follows
\begin{equation}
  \sum_{j=1,3} \nabla^2_j =
   \biggl [ {\partial^2 \over \partial\rho^2}
  +{8\over \rho} {\partial \over \partial\rho} +{\Lambda^2(\Omega_p)
  \over \rho^2}\biggr]\ , \label{eq:kin}
\end{equation}
and
\begin{equation}\label{eq:kin2}
   \biggl ( \Lambda^2(\Omega_p) + K(K+7) \biggr)
      {\cal Y}^{K,LM}_{\ell_1,\ell_2,\ell_3, L_2 ,n_2, n_3}(\Omega_p)
      = 0 \ .
\end{equation}
Another important property of the HH functions is that $\rho^K   {\cal
Y}^{K,L,M}_{\ell_1,\ell_2,\ell_3, L_2 ,n_2, n_3}(\Omega_p)$ are
homogeneous polynomials of the particle coordinates of degree $K$.

The  WF of a state with total angular momentum $J$, parity $\pi$ and
total
isospin $T$ can be expanded over the following complete basis of
antisymmetrical hyperangular--spin--isospin states, defined as
\begin{equation}
  \Psi_{\mu}^{KLSTJ\pi} = \sum_{p=1}^{12}
  \Phi_\mu^{KLSTJ\pi}(i,j;k;m)\ ,
  \label{eq:PSI}
\end{equation}
where the sum is over the $12$ even permutations $p$, and
\begin{equation}
  \Phi^{KLSTJ\pi}_{\mu}(i,j;k;m) =  \biggl \{
   {\cal Y}^{K,LM}_{\ell_1,\ell_2,\ell_3, L_2 ,n_2, n_3}(\Omega_p)
      \biggl [\Bigl[\bigl[ s_i s_j \bigr]_{S_a}
      s_k\Bigr]_{S_b} s_m  \biggr]_{S} \biggr \}_{JJ_z}
      \biggl [\Bigl[\bigl[ t_i t_j \bigr]_{T_a}
      t_k\Bigr]_{T_b} t_m  \biggr]_{TT_z}\ .
     \label{eq:PHI}
\end{equation}
Here, ${\cal Y}^{KLM}_{\ell_1,\ell_2,\ell_3, L_2 ,n_2,
n_3}(\Omega_p)$ is the HH state defined in Eq.~(\ref{eq:hh4}), and
$s_i$ ($t_i$) denotes the spin  (isospin) function of particle $i$.
The total orbital angular  momentum $L$ of the HH function is
coupled to the total spin $S$ to give a total angular momentum $J$,
$J_z$. The quantum number $T$ specifies the total isospin, while
$\pi=(-1)^{\ell_1+\ell_2+\ell_3} $ is the parity of the state.
The integer index $\mu$ labels the possible choices of hyperangular,
spin and isospin quantum numbers, namely
\begin{equation}
   \mu \equiv \{ \ell_1,\ell_2,\ell_3, L_2 ,n_2, n_3, S_a,S_b, T_a,T_b
   \}\ ,\label{eq:mu}
\end{equation}
compatible with the given values of $K$, $L$, $S$, $T$, $J$ and $\pi$.
Another important classification of the states is to group them into
``channels'': states belonging to the same channel have the same
values of angular $\ell_1,\ell_2,\ell_3, L_2 ,L$, spin $S_a,S_b,S$ and
isospin $T_a,T_b,T$ quantum numbers but different values of $n_2$, $n_3$.

Each state  $\Psi^{KLSTJ\pi}_\mu$ entering the expansion of the
four-nucleon WF must to be antisymmetric under the exchange of any
pair of particles.
Consequently, it is necessary to consider states such that
\begin{equation}
    \Phi^{KLSTJ\pi}_\mu(i,j;k;m)= -\Phi^{KLSTJ\pi}_\mu(j,i;k;m)\ .
     \label{eq:exij}
\end{equation}
Under the exchange  $i \leftrightarrow j$, the Jacobi vector $\jacb_{3p}$
changes its sign, whereas $\jacb_{1p}$ and $\jacb_{2p}$ remain
unchanged, and, therefore, the HH function ${\cal
Y}^{KLM}_{\ell_1,\ell_2,\ell_3, L_2 ,n_2,n_3}(\Omega_p)$ transforms into
itself times a factor $(-1)^{\ell_3}$
(see Eqs.~(\ref{eq:JcbV}) and~(\ref{eq:hh4})). On the other hand, the
spin--isospin part  transforms into
itself times a factor $(-1)^{S_a+T_a}$ for the $i \leftrightarrow j$ exchange.
Thus,  the condition~(\ref{eq:exij})
is fulfilled when
\begin{equation}
    \ell_3+S_a+T_a = {\rm odd}\ . \label{eq:lsa}
\end{equation}

The number $M_{KLSJT\pi}$ of the antisymmetrical functions
$\Psi^{KLSJT\pi}_\mu$ having given  $K$, $L$, $S$,
$T$, $J$ and $\pi$ values but different combination of
the quantum numbers $\mu$ is in general very large.  In addition to the
degeneracy $N_{KL\pi}$ of the HH basis, the four spins (isospins) can
be coupled in different ways to $S$ ($T$).
However, many of the states $\Psi^{KLSJT\pi}_\mu$
are linearly dependent
amongst themselves. In the expansion of a four-nucleon WF
it is necessary to include the subset of linearly independent states only.
To search for  the independent states, the essential ingredient is
the knowledge of
\begin{equation}
   N^{KLSTJ\pi}_{\mu\mu'}=\langle \Psi^{KLSTJ\pi}_\mu |
                    \Psi^{KLSTJ\pi}_{\mu'} \rangle_\Omega
   \ ,\label{eq:norma}
\end{equation}
where $\langle \rangle_\Omega$ denotes the evaluation of the
spin--isospin traces and the integration over the hyperspherical
variables. 

The calculation of the matrix elements of the Hamiltonian is
considerably simplified by using the following transformation
\begin{equation}\label{eq:arare}
  \Phi^{KLSTJ\pi}_{\mu}(i,j;k;m) =
  \sum_{\mu'}  a^{KLSTJ\pi}_{\mu,\mu'}(p) \Phi^{KLSTJ\pi}_{\mu'}(1,2;3;4)\ .
\end{equation}
The coefficients $a^{KLSTJ\pi}_{\mu,\mu'}(p)$ have been obtained using
the techniques described in Ref.~\cite{V98}.
The  states $\Psi_{\mu}^{KLSTJ\pi}$ can be written as
\begin{equation}
  \Psi_{\mu}^{KLSTJ\pi} =
  \sum_{\mu'} A^{KLSTJ\pi}_{\mu,\mu'} \Phi^{KLSTJ\pi}_{\mu'}(1,2;3;4)\ ,
  \label{eq:PSI2}
\end{equation}
where
\begin{equation}\label{eq:arare2}
  A^{KLSTJ\pi}_{\mu,\mu'} =  \sum_{p=1}^{12} a^{KLSTJ\pi}_{\mu,\mu'}(p)\ .
\end{equation}
The  matrix elements of the norm can be easily
obtained using the orthonormalization of the HH basis
with the result that:
\begin{equation}
  N^{KLSTJ\pi}_{\mu\mu'}=
  \sum_{\mu''} ( A^{KLSTJ\pi}_{\mu,\mu''})^*
   A^{KLSTJ\pi}_{\mu',\mu''}\ .
  \label{eq:normaB}
\end{equation}
Clearly,
\begin{equation}
   \langle \Psi^{KLSTJ\pi}_\mu |
                    \Psi^{K'L'S'T'J'\pi'}_{\mu'} \rangle_\Omega=0
   \ ,\qquad {\rm if\ }\{KLSTJ\pi\}\ne\{K'L'S'T'J'\pi'\}\ .
   \label{eq:normaC}
\end{equation}
Once the quantities $N^{KLSTJ\pi}_{\mu\mu'}$ are calculated, the
Gram--Schmidt procedure can be used, for example, to eliminate the
linear dependent states between the various $\Psi^{KLSTJ\pi}_\mu$
functions.

We have found that the number of independent states $M'_{KLSTJ\pi}$
for given $K$, $L$, $S$, $T$, $J$ and $\pi$ is noticeably
smaller than the corresponding value of $M_{KLSTJ\pi}$ . To give an
example, we have reported in Table~\ref{tb:nhh} a few values of
$M_{KLSTJ\pi}$ and $M'_{KLSTJ\pi}$ for the case $J=0$, $T=0$,
$\pi=+$ corresponding to the ground state of the $\alpha$-particle.
As can be seen from the table, the values of $M_{KLL00+}$
are very large also for moderate values of $K$, but
$M^\prime_{KLL00+}$ are usually much smaller.

The total WF can finally be written as
\begin{equation}\label{eq:PSI3}
  \Psi^{J\pi}_4= \sum_{KLST}\sum_{\mu} {u_{KLST,\mu}(\rho)\over \rho^4}
  \Psi_{\mu}^{KLSTJ\pi}\ ,
\end{equation}
where the sum is restricted only to the linearly independent
states. The expansion coefficients, which depend on the
hyperradius, are determined by the Rayleigh--Ritz variational
principle. By applying this principle, a set of second order
differential equations for the functions $u(\rho)$ are obtained.
These equations and the procedure adopted to solve them has
been outlined in the appendix of Ref.~\cite{KMRV96}. In this way,
a large number of equation can be solved.

The main problem is the computation of the matrix elements of
the Hamiltonian. The kinetic energy operator matrix elements are
readily calculated analytically, whereas the  matrix
elements of a local NN potential can be obtained by one dimensional 
integrations. To this aim, it is convenient to write the basis in the 
jj coupling scheme
\begin{equation}
  \Psi_{\mu}^{KLSTJ\pi} =
  \sum_\nu B^{KLSTJ\pi}_{\mu,\nu} \; \Xi^{KTJ\pi}_{\nu}(1,2;3;4)\ ,
  \label{eq:PSI3jj}
\end{equation}
where
\begin{eqnarray}
  \Xi^{KTJ\pi}_{\nu}(1,2;3;4) &=&  \biggl \{
   \left [ \Bigl ( Y_{\ell_3}(\widehat \jac_{3}) 
          ( s_1 s_2 )_{S_a} \Bigr)_{j_3}
           \Bigl ( Y_{\ell_2}(\widehat \jac_{2}) s_3 \Bigr)_{j_2}
   \right]_{J_2}
          \Bigl ( Y_{\ell_1}(\widehat \jac_{1}) s_4
   \Bigr)_{j_1}\biggr\}_{JJ_z}\times
      \nonumber \\
  &&    \times \biggl [\Bigl[\bigl[ t_i t_j \bigr]_{T_a}
      t_k\Bigr]_{T_b} t_m  \biggr]_{TT_z} \;
     {\cal P}^{\ell_1,\ell_2,\ell_3}_{n_2, n_3}(\hypfi_{2p},\hypfi_{3p})\ ,
     \label{eq:PHIjj}
\end{eqnarray}
and $B^{KLSTJ\pi}_{\mu,\nu}$ are related to the
coefficients $A^{KLSTJ\pi}_{\mu,\mu'}$ via Wigner 3j and 6j coefficients.
Now, the integer index $\nu$ labels all possible choices of
\begin{equation}
   \nu\equiv\{n_3,\ell_3,S_a,j_3,n_2,\ell_2,j_2,J_2,\ell_1,j_1,T_a,T_b\}
   \ ,\label{eq:nu}
\end{equation}
compatible with the given values of $K$, $T$, $J$ and $\pi$.

In terms of the states $\Xi^{KTJ\pi}_{\nu}(1,2;3;4)$, it is easy to
compute the matrix elements of an NN potential. For example, the matrix
element of the isospin-conserving part $V_{IC}(1,2)$ of the NN potential
\begin{equation}
 \langle \Xi^{KTJ\pi}_{\nu}(1,2;3;4) | V_{IC}(1,2) |
  \Xi^{K'T'J\pi}_{\nu'}(1,2;3;4)\rangle_\Omega=0\ ,
  \label{eq:nnpi1}
\end{equation}
unless
\begin{equation}
 \{j_3,n_2,\ell_2,j_2,J_2,\ell_1,j_1,T_a,T_b,T\}=
 \{j_3',n_2',\ell_2',j_2',J_2',\ell_1',j_1',T_a',T_b',T'\}\ .
 \label{eq:nnpi2}
\end{equation}
If Eq.~(\ref{eq:nnpi2}) is verified, then
\begin{eqnarray}
 \lefteqn{ \langle \Xi^{KTJ\pi}_{\nu}(1,2;3;4) | V_{IC}(1,2) |
  \Xi^{K'TJ\pi}_{\nu'}(1,2;3;4)\rangle_\Omega=}\qquad\qquad
   \qquad &&  \nonumber \\
 &=& {\cal N}^{\ell_1,\ell_2,\ell_3}_{n_2,n_3}
    {\cal N}^{\ell_1,\ell_2,\ell_3'}_{n_2,n_3'}
    \int_0^{\pi\over 2} d\hypfi_3 \; (\cos\hypfi_3)^{2+\ell_3+\ell_3'}
    ( \sin\hypfi_3)^{5+2\ell_1+2\ell_2+4 n_2} \times \nonumber \\
 &&   \qquad\times \;
    v^{j_3}_{\ell_3,S_a,\ell_3',S_a'}(\rho\cos\hypfi_3)\;
    P^{\ell_1+\ell_2+2n_2+2, \ell_3+{1\over
       2}}_{n_3}(\cos2\hypfi_{3}) \times\nonumber \\
  && \qquad\times
      P^{\ell_1+\ell_2+2n_2+2, \ell_3'+{1\over
      2}}_{n_3'}(\cos2\hypfi_{3})\ , \label{eq:nnpi3}
\end{eqnarray}
where $v^{j}_{\ell,S,\ell',S'}(r)$ is the isospin-conserving part of the
NN potential acting between two-body states ${}^{2S+1}(\ell)_j$
and ${}^{2S'+1}(\ell')_j$. The one-dimensional integral
given in Eq.~(\ref{eq:nnpi2}) can be computed numerically with
high accuracy. The case of the isospin-breaking part of the NN
interaction is a generalization of the previous case: now we can
have $\{T_a,T_b,T\}\neq \{T_a',T_b',T'\}$ as well.

The 3N interaction matrix elements are more difficult to compute and
the adopted procedure is detailed  in the Appendix.

\section{Choice of the basis}
\label{sec:bas}

The main difficulty of applying the HH technique is the selection
of a restricted and effective subset of basis states. In fact,
although the number of independent states proves to be much smaller than the
degeneracy $M_{KLSTJ\pi}$ of the basis, the brute force application of
the method, i.e., the inclusion of all HH states 
having $K\le K_{\rm M}$ in the expansion and then
increasing $K_{\rm M}$ until convergence, would be destined 
to fail. In fact, due to the strong
correlations induced by the NN potential, $K_{\rm M}\approx 60$ are 
necessary in order to obtain a good convergence. However, already for values 
of $K> 20$ it is very difficult to find the linearly independent states 
via the Gram-Schmidt procedure due to the loss of precision in the 
orthogonalization procedure.

However, it is possible to separate the HH functions into classes having
particular properties and  advantageously take into account the fact that
the convergence rates of the various classes are rather different.  As
discussed in the Introduction, we expect that the contribution
of the HH functions describing the two-body correlations to be very
important~\cite{F83}. Another criterion adopted is first to consider
the HH functions with low values of $\ell_i$. 

An important quantity in the choice of the classes 
is ${\cal M}_i(K_i)$, namely the number of linearly independent
antisymmetrical spin-isospin-HH states $\Psi_{\mu}^{KLSTJ\pi}$
belonging to a class $i$ and having $K\le K_i$. 
Only even parity states have been included
in the construction of the $\alpha$-particle WF, and thus
the discussion hereafter will be limited to consider only even values  
for $K_i$. In general, for a class $i$, the value
${\cal M}_i(K_i)$ is zero for $K_i < K_i^a$, due to the fact that
the linearly-dependent states have been removed from the expansion.
For example, as should be clear by inspection of Table~\ref{tb:nhh},
there is only one linearly-independent state $\Psi_{\mu}^{KLS00+}$
with $K=0$. If this state is included in the first class, the other 
classes must have at least $K_i^a=2$, etc. For $K_i\gg K_i^a$,
${\cal M}_i(K_i)$ reaches a sort of ``asymptotic'' value,
given by ${\cal M}_i(K_i)\approx M_i (K_i)^{r_i}$.
The choice of the classes has clearly to be optimized so that 
the convergence for the classes with large values of $M_i$ and $r_i$
could be reached for relatively low values of $K$.
The specific values of $M_i$ and $r_i$ are discussed below.

To study the $\alpha$-particle ground state we have found it very
convenient to choose as follows ($T$ is the total isospin):
\begin{enumerate}
  \item {Class C1}. In this class the $T=0$ HH states 
   belonging to the PB are included . For $A=4$,
   the PB includes states
   of the first three channels reported in Table~\ref{tb:chan0p0} (the
   only channels with $\ell_1=\ell_2=0$) with $n_2=0$. As can be seen from
   Eq.~(\ref{eq:hh4}), the corresponding states depends only on $\widehat
   \jacb_{3p}$ and $\cos\hypfi_{3p}=\jac_{3p}/\rho\equiv r_{ij}/\rho$, 
   and therefore contain only two-body correlations. For this class,
   $K_1^a=0$. For $K_1\ge 4$, ${\cal M}_1(K_1)=
   (3/2) K_1$. 
   Then, this is a ``small'' class. As will be shown in the next
   Section, this is also the most slowly convergent class,
   but since  $M_1=(3/2)$ and $r_1=1$,
   it is not difficult to reach the
   desired degree of accuracy. 

  \item {Class C2}. This class includes the $T=0$ states belonging to the
  same three channels as those of class C1, but with $n_2>0$. These states
  therefore include also part of the three--body correlations.
  The first linearly-independent states of this class
  appear for $K=4$, therefore $K_2^a=4$. Moreover,
  ${\cal M}_2(K_2)=(3/4) (K_2)^2
  +{\cal O}(K_2)$ for $K_2\gg 1$. This can be considered
  a ``small'' class, too, and
  states up to $K_2=40$ have been included in the
  present calculation without difficulty.

  \item {Class C3}. This class includes the remaining $T=0$ states
  of the channels having $\ell_1+\ell_2+\ell_3=2$.  The corresponding 20
  possible channels are reported in Table~\ref{tb:chan0p0} in rows $4-23$. 
  In this case
  $K_3^a=2$ and  ${\cal M}_3(K_3)=5 (K_3)^2+{\cal O}(K_3)$ for $K_3\gg 1$.
  This is a fairly ``large'' class,
  but with the necessary care states with
  $K_3\approx 34$ can be still included in the expansion.

  \item {Class C4}. This class includes $T=0$  states belonging
  to the channels with $\ell_1+\ell_2+\ell_3=4$. There are
  57 channels of this kind. In this case
  $K_4^a=8$ and it follows that 
  ${\cal M}_4(K_4)=(57/4) (K_4)^2+{\cal O}(K_4)$ for $K_4\gg 1$. This 
  is a ``large'' class, but its contribution to the $\alpha$-particle
  BE is, though still sizable,  far less important than the
  first three classes. States of up to 
  $K_4\approx 28$ have been considered.

  \item {Class C5}. This class includes $T=0$  states belonging
  to the channels with $\ell_1+\ell_2+\ell_3=6$. There are
  109 channels of this kind. In this case
  $K_5^a=12$ and for $K_5\gg 1$ we have
  ${\cal M}_5(K_5)=(109/4) (K_5)^2+{\cal O}(K_5)$. This 
  is a very ``large'' class, but it contributes
  very little to the $\alpha$-particle BE, as we shall see.
  Therefore, we can truncate the expansion already at $K_5\approx 20$.

  \item {Class C6}. This class includes the states having $T>0$. We
  have included in the expansion all the channels of this kind with
  $\ell_1+\ell_2+\ell_3 \le 2$ (45 channels). In this case
  $K_6^a=0$ and for $K_6\gg 1$ we have
  ${\cal M}_6(K_6)=(45/4) (K_6)^2+{\cal O}(K_6)$. 
  Also the contribution to this class is very tiny.

\end{enumerate}

The states belonging to the classes C2 and C3 describe the most important
three-body contributions to the WF. The classes C4 and C5 take into
account the remaining three and four body correlations ordered with increasing
values of $\ell_1+\ell_2+\ell_3$. 

The convergence is studied as follows. First, only the states
of class C1 with $K\le K_1$ are included in the expansion and the
convergence of the BE is
studied as the value of $K_1$ is increased. Once a satisfactory value of
$K_1=K_{1\rm M}$ is reached, the states of the second class with $K\le
K_2$ are added to the expansion, keeping all the states of the class
C1 with $K\le K_{1\rm M}$. Then $K_2$ is increased up to $K_{2M}$
in order to reach the desired convergence for the BE. With some extra work, 
it is possible at this point to optimize the basis by removing some
of the $K\le K_{2M}$ states of class C2 which give
very tiny contributions to the BE.
The procedure outlined is then repeated for each new class. 
Our complete calculation includes about $8,000$ states.

It should be noticed that in the present calculation only HH functions
constructed in terms of the Jacobi vectors given in
Eq.~(\ref{eq:JcbV}), referred to as the set A,  have been considered. As
is well known, there is another possible choice, namely
\begin{eqnarray}
          \jacb'_{1p}& = & \bfr_m-
                             \bfr_k \ , \nonumber\\
         \jacb'_{2p} & = & \sqrt{1\over2} (\bfr_k+\bfr_m -
              \bfr_i-\bfr_j )\ ,  \label{eq:JcbV2}\\
         \jacb'_{3p} & =& \bfr_j-\bfr_i\ , \nonumber
\end{eqnarray}
hereafter referred as the set B of Jacobi vectors.
Considering, for example, the $\alpha$-particle ground state,
the HH functions ${\cal Y}_{\rm set\ A}$ of the set A 
are more appropriate for describing those contributions
to the WF corresponding to $\{3+1\}$ clustering structures, namely
$\hel+n$ or $\tri+p$. The HH functions ${\cal Y}_{\rm set\ B}$
constructed with the set B, should be more suitable for describing
the $\{2+2\}$ clustering structures, such as the $d+d$ configurations. 
It is rather obvious that the inclusion of HH functions of 
both sets should speed up the convergence in constructing the
full state of the system~\cite{KK90,KRV95}. If the 
expansion of the WF is done over only a particular set, 
those configurations in which other clustering structures are important 
would be generally described with difficulty and a slow convergence 
would result.

In the present calculation we have included HH states 
${\cal Y}_{\rm set\ A}$ only, i.e. constructed with the Jacobi vectors
of the set A, since this has been found to be sufficient 
to reach the desired degree of convergence.
In fact, the full basis considered (classes C1-C6) is large
enough to include all the possible independent states for $K\le
20$. Additional linearly independent states
constructed with the set B would appear only for $K\ge 22$. As  will
appear clear below, the contribution of states with $K\ge 22$ {\it
not belonging} to classes C1--C3 is rather small. Therefore, in the
present calculation it is not necessary to introduce states of the set
B. However in the present formalism there would be no
particular difficulty in also including states constructed with the 
set B.

\section{Results for the $\alpha$ particle ground state}
\label{sec:res}

In this section, the results obtained for the ground state of the
$\alpha$ particle are presented.
The convergence of the HH expansion in terms of the
various classes is examined in
Sect.~\ref{sec:conv}. The results obtained for the BE
and other properties for a number of
different interaction models  are reported in Sect.~\ref{sec:resv}.
The origin of the $T>0$ components in the $\alpha$-particle ground state
is discussed in Sect.~\ref{sec:cib}. The effect of
the truncation of the NN and 3N interactions is studied
in Sect.~\ref{sec:trun}. Finally, the calculation of the various
$\heq$ asymptotic normalization constants is considered in 
Sect.~\ref{sec:anc}.

\subsection{Convergence}
\label{sec:conv}

In order to study the convergence,
we have considered three different interaction models frequently
used in literature.
The first calculation has been performed using the
MT-V potential~\cite{MT69}, a central spin-independent interaction.
The parameters defining this potential can be found in Table~I
of Ref.~\cite{VS95} and we have used $\hbar^2/m=41.47$ MeV fm$^2$.
This potential has been used for a number of benchmarks. It does 
not contain any non central components, but it retains a rather 
strong repulsion at short interparticle distance going like $1/r$. 
It is therefore rather challenging for a technique where the correlations 
are not built in.
In the second example, we have considered the
AV18 potential model~\cite{AV18}  which represent
a NN interaction in its full richness, with short-range repulsion,
tensor and other non-central components, charge symmetry breaking
terms, and Coulomb and other electromagnetic interactions. In the
third case,
we have added to the AV18 potential the Urbana IX
model~\cite{AV18+} of 3N interaction (AV18+UIX model). For the latter
two models we have used $\hbar^2/m=41.47108$ MeV fm$^2$.

We study the convergence as explained in the previous Section, and
the results presented in table~\ref{tb:conv} are arranged accordingly.
For example, the BE $B$ reported in a row with a given set of values of
$K_1,\ldots,K_6$ has been obtained by including in the expansion all
the HH functions of class ${\rm C}_i$ with $K\le K_i$, $i=1,\ldots,6$.

For the MT-V potential,
we observe a slow convergence of the classes C1 and C2
and fairly large values of $K$ have to be used.
On the other side, they give 96\% of the total BE.
The contributions of the other classes are extremely small. The class C3
increases the BE by an additional $0.08$ MeV, and the class C4 by
less than $0.01$ MeV. Class C5 gives a negligible contribution, and
class C6 has not been included in the expansion since for this potential
isospin is a good quantum number and there is not any mixing with $T>0$
components. The final value $B=31.347$ MeV is in good agreement with the
results found in literature, whose ``average'' has been
reported in the last row of Table~\ref{tb:conv}. There is
approximately 10 keV of missing energy  due to the truncation
of our expansion as will be discussed at the end of this Subsection.

For the AV18 potential, the first two classes give important contributions
but a large amount of BE is still missing.
The inclusion of the third class increases the BE by more than $3$ MeV
but $0.8$ MeV are still missing. Since the second and third
classes take into account a large part of the contributions of the
three body correlations, this means that
also the four body correlation are important. These are related to the
configurations where the clusterization $2+2$ is important. In our
calculation, such configurations are included when the classes C4 and C5
are taken into account. The number of the states of class C4 increases
very rapidly with $K_4$ but fortunately the convergence
is reached around $K=24$. The gained BE is almost
$0.8$ MeV.  There are no linearly independent states of class C5 
with $K<14$ and its contribution is rather small. The
convergence is again obtained around $K=24$, but the gain in energy is
only about $0.02$ MeV. Since the number of states of this class
is very large, for example  ${\cal M}_6(20)\approx 800$ when confronted
with a very tiny gain in BE, a selection of the states has 
to be performed to save computing time and to avoid loss of
numerical precision. For example, all the channels 
of class C5 with a total orbital angular momentum $L=0$ have not been 
included in the expansion since their contribution is absolutely 
negligible. With their inclusion the procedure of
Ref.~\cite{KMRV96} for finding the eigenvalue would become 
numerically instable.

From table~\ref{tb:conv},
one can try to estimate the contribution of the states  with
${\cal L}=\ell_1+\ell_2+\ell_3=8$. From the previous discussion
we have already seen that the states having ${\cal L}=4$ (class C4)
contribute by about $0.8$ MeV, while the states with ${\cal L}=6$
(class C5) contribute by
less than  $0.03$ MeV. Therefore, the states with ${\cal L}=8$
are expected to give a negligible contribution to the $\a$-particle
BE. Finally, the inclusion of states with $T>0$ (class
C6) increases the BE by another  20 keV, approximately.

The convergence rate when considering the UIX 3N interaction is
similar to the AV18 case. The corresponding results
are reported in the last column of table~\ref{tb:conv} 
(they have been obtained in the approximation described in 
Subsec.~\ref{sec:trun}). Since the models most frequently used for the 3N 
interactions are rather soft at short interparticle distances,
the convergence rate of the C1 and C2 classes does not change
appreciably. However, the 3N potential has a very strong state dependence
and the convergence of the C3-C5 classes are now slightly slower.
For example, the gain in BE of the C4 class is about $0.8$ MeV without
any 3N interactions, and  it becomes about $1$ MeV when including the 3N
interaction.
Our final results for the AV18 and AV18+UIX models agree well with the
FY results of Ref.~\cite{Nea02} reported in the last row of
Table~\ref{tb:conv}. 
The convergence properties for other NN and NN+3N potential models
has been found  rather similar to those showed in
Table~\ref{tb:conv}.

Finally, let us comment about the convergence rate of the
expansion as a function of the maximal grand angular
quantum numbers $K_i$ of the various classes of HH states
included in our expansion. Previous studies~\cite{ZPE69,S72,D77,F83}
have shown that 
the trend of convergence toward the exact BE depends
primarily on the form of the potential. In particular, for potentials
which are given as functions of $r_{ij}^2$ (as, for example, those given as 
a sum of gaussians) the increase of BE with $K_i$ diminishes
exponentially. On the other hand, for potentials given as a function of
$r_{ij}$ (as a sum of exponentials or Yukawians), the increase of
BE decreases as $(1/K_m)^p$, where $p$ is a positive
integer number. The value of $p$ is smaller for potentials of Yukawa
type due to the $1/r$ divergence at the origin, but may depends also
on the class of the HH functions whose convergence is
studied. It is important to determine the value of
$K_m$ at which the convergence starts to behave as stated previously.
The asymptotic behavior of
the convergence should be reached for HH functions whose kinetic energy
$\propto (\hbar^2/m) K(K+7)/\rho_0^2$ is much greater than the BE,
where $\rho_0$ is a value of the hyperradius $\rho$ for
which $\Psi(\rho_0)$ can be regarded as small~\cite{D77}.
In our studies, we have found that the asymptotic falling begins
for $K_m\approx 30\div 40$.

\begin{figure}[p]
\includegraphics[scale=.9]{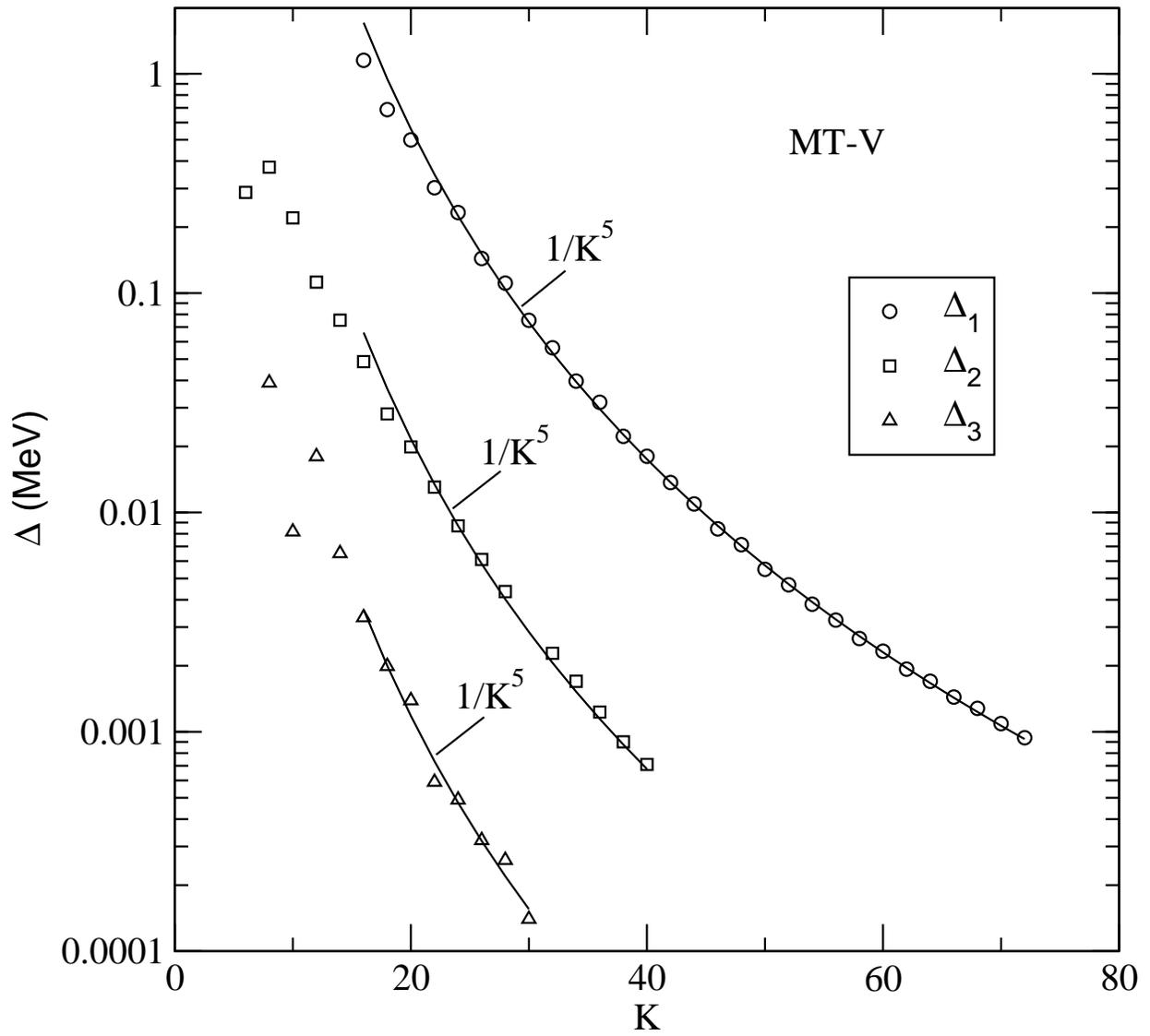}
\caption{Binding energy differences for the $\alpha$ particle
for the classes C1 (circles) C2 (squares) and C3 (up triangles)
as function of the grand angular value $K$ (see the text for more
details). The potential used is the MT-V. The curves are
fitted to the large $K$ part of  
the energy differences.}
\label{fig:diff-MT}
\end{figure}

In order to study the convergence behavior we have indicated with
$B(K_1,K_2,K_3,K_4,K_5,K_6)$ the BE obtained
by including in the expansion all the HH states of the class C1 with $K\le
K_1$, all the HH states of the class C2 having $K\le K_2$, etc.
Let us compute
\begin{eqnarray}
  \Delta_{1}(K)&=&B(K,0,0,0,0,0)-B(K-2,0,0,0,0,0)\ ,
  \label{eq:c1diffa}\\
  \Delta_{2}(K)&=&B(K_{1{\rm M}},K,0,0,0,0)-
                  B(K_{1{\rm M}},K-2,0,0,0,0)\ ,\qquad K_{1{\rm M}}=72\ ,
      \label{eq:c2diff}\\
  \Delta_{3}(K)&=&B(K_{1{\rm M}},K_{2{\rm M}},K,0,0,0)-
                  B(K_{1{\rm M}},K_{2{\rm M}},K-2,0,0,0)\ ,
                  \qquad K_{1{\rm M}}=72\ ,K_{2{\rm M}}=40\ ,
  \label{eq:c3diff}
\end{eqnarray}
and so on. The values obtained for $\Delta_{i}$, $i=1$, $3$ are shown in
Fig.~\ref{fig:diff-MT} for the MT-V potential model, together
with the curves $(1/K)^p$ for $p=5$ (the curves
have been constrained to fit the high $K$ part of the
$\Delta_{1\div3}(K)$ values ). As can be seen in Fig.~\ref{fig:diff-MT},
all the  energy differences $\Delta_1$, $\Delta_2$ and
$\Delta_3$ decrease as $1/K^5$ for $K\ge  20$, approximately. However,
for a given $K$, there is a clear hierarchy
$\Delta_1(K)\gg\Delta_2(K)\gg\Delta_3(K)$. Note that
there are slight 
fluctuations in the $\Delta(K)$ as $K$ is increased (this is evident
in particular for $\Delta_3$).

\begin{figure}[p]
\includegraphics[scale=.9]{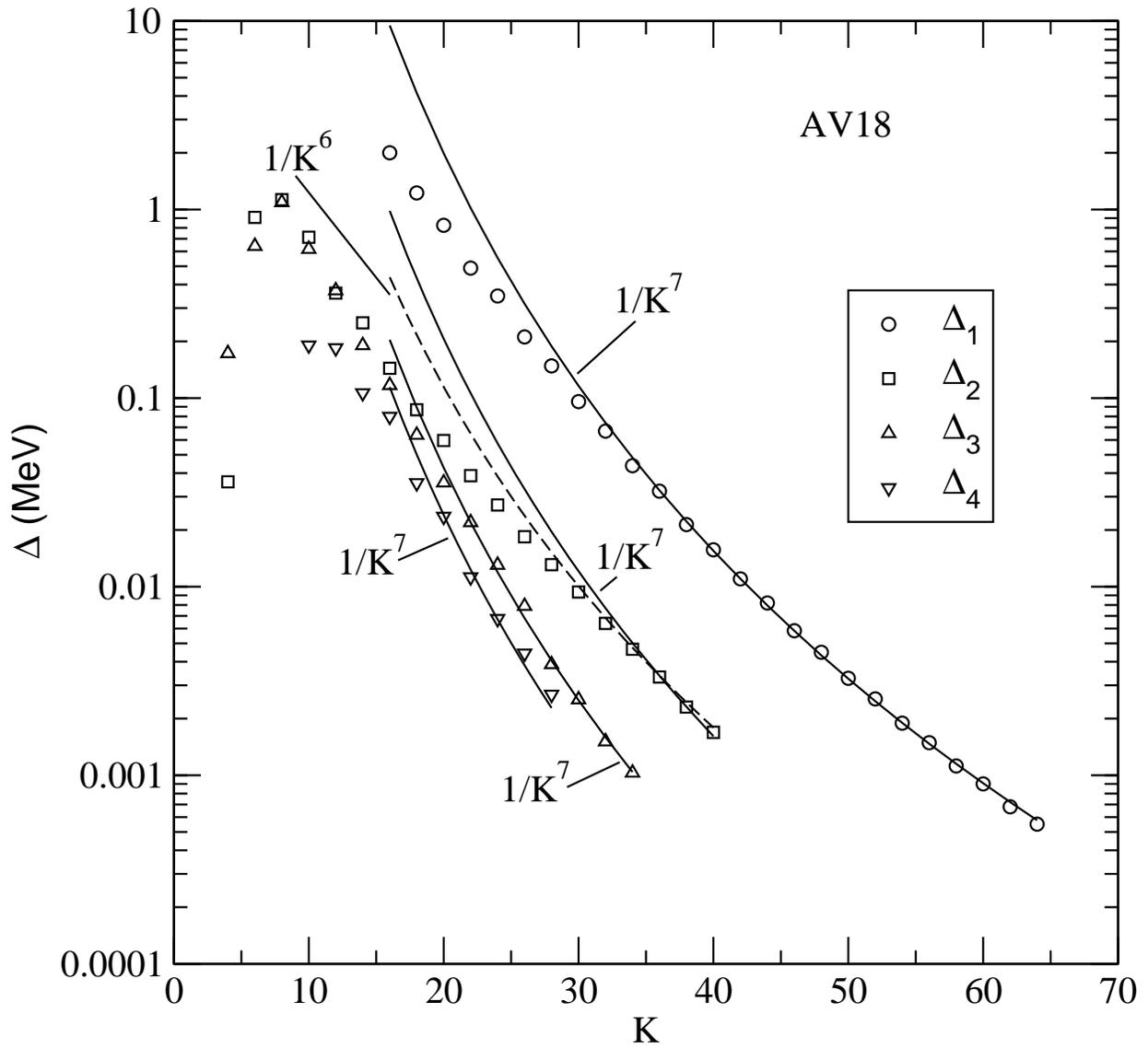}
\caption{Binding energy differences for the $\alpha$ particles
for the classes C1 (circles) C2 (squares), C3 (up triangles) and C4
(down triangles) as function of the grand angular value $K$ (see the
text for more details). The potential used is the AV18. The curves are
fitted to the large $K$ part of   
the energy differences.}
\label{fig:diff-AV18}
\end{figure}

The values obtained for $\Delta_{i}$, $i=1$,$4$ for the AV18 potential
are reported in Fig.~\ref{fig:diff-AV18}. The decrease of
$\Delta_1$, $\Delta_3$ and $\Delta_4$ clearly follows a law
$(1/K)^p$ with $p=7$. The behavior of the energy difference
$\Delta_2$ can be approximated either by a $1/K^6$ or a $1/K^7$ law. 
The faster decrease of these results compared to the previous case is 
due to the fact that the AV18 potential does not diverge at the origin, 
while the MT-V has a $1/r$ divergence. The study of
Ref.~\cite{D77}, in fact,  predicts a difference of two units in
the exponential coefficient for the two cases (Yukawian potentials
vs. regular potentials). For fixed $K$, also for the AV18 we note a
systematic hierarchy $\Delta_1(K)\gg\Delta_2(K)\gg\Delta_3(K)\gg
\Delta_4(K)$, although less pronounced than in the MT-V case.
The same behaviour is observed when the UIX 3N potential is
included.

From the observed simple behavior, we can readily estimate the
missing BE  due to the truncation of the expansion to finite values of
$K=\overline{K}$.
Let us suppose that the states of class $i$ up to
$K=\overline{K}$ have been included and to have computed 
$\Delta_i(\overline{K})$. Then, the missing BE due to the states with
$K=\overline{K}+2$, $\overline{ K}+4$, $\ldots$, is given by
\begin{equation}
   (\Delta B)_i =  c(\overline{ K},p) \; \Delta_i(\overline{ K})\ ,
   \qquad  c(\overline{ K},p)= \sum_{K=
   \overline{ K}+2,\overline{K}+4,\ldots}^\infty  \left(
    {\overline{ K}\over K} \right )^p \ ,
   \label{eq:extra}
\end{equation}
where $c(\overline{ K},p)$ is a numerical coefficient.
For example, let us consider the ``missing'' energy for the Class C1
in the MT-V case. In this case $\overline{ K}=72$ and $p=5$ and
$c(72,5)=8.51$. Since $\Delta_1(\overline{ K}=72)=0.99$ keV, we find 
that $\Delta B_1=8.4$ keV. Adding this value to $B(72,0,0,0,0,0)$
we can extrapolate the BE for the case of the inclusion of the {\it whole}
class C1: $B(72,0,0,0,0,0)+(\Delta B)_1\approx 30.041$ MeV.
The BE obtained corresponding to this case (converged PB expansion)
has been computed very precisely in Ref.~\cite{VKR92}
using a ``pair-correlated'' potential basis (PPB). 
In such an expansion, each PB function is multiplied by a pair
correlation factor and this allows for a very rapid convergence of the
expansion. In that paper, we
found  $B({\rm PPB})=30.042$ MeV which agrees very well with the
above extrapolated value. To reach such a value, it would be necessary
to use $\overline{ K}\approx 150$ for the class C1.

For the AV18, we find $c(72,7)=5.52$, $\Delta_{1}(\overline{ 
K}=72)=0.24$  keV and therefore $(\Delta B)_1\approx 1.3$
keV, a rather tiny quantity. Due to the faster convergence for this
potential like $1/{ K}^7$, it does not seem necessary to
increase $K_1$ any further  in this case.

The ``missing'' energy of the other classes can be estimated in the
same way. However, to estimate the ``missing'' energy for the whole
calculation due to the truncation of the expansion of the
first class up to $K\le K_1$, of the second class up to $K\le K_2$,
etc., we cannot simply add the $(\Delta B)_i$,
$i=1,\ldots,6$ so obtained. The reason is that, for example, the
inclusion of the HH states of classes C2, C3, $\ldots$,  also
alters the convergence of class C1, etc. by a small amount. To study the
``full'' rate of convergence, let us consider 
\begin{eqnarray}
  \bar\Delta_{1}(K)&=&B(K,K_{2{\rm M}},K_{3{\rm M}},K_{4{\rm
  M}},K_{5{\rm M}},K_{6{\rm M}})- \nonumber \\
   && \qquad\qquad B(K-2,K_{2{\rm M}},K_{3{\rm
  M}},K_{4{\rm M}},K_{5{\rm M}},K_{6{\rm M}})\ , 
  \nonumber\\
  \bar\Delta_{2}(K)&=&B(K_{1{\rm M}},K,K_{3{\rm M}},K_{4{\rm
  M}},K_{5{\rm M}},K_{6{\rm M}})-\nonumber \\
   && \qquad\qquad  B(K_{1{\rm M}},K-2,K_{3{\rm
  M}},K_{4{\rm M}},K_{5{\rm M}},K_{6{\rm M}})\ , 
      \label{eq:c2diffb}\\
  \bar\Delta_{3}(K)&=&B(K_{1{\rm M}},K_{2{\rm M}},K,K_{4{\rm
  M}},K_{5{\rm M}},K_{6{\rm M}})-\nonumber \\
   && \qquad\qquad B(K_{1{\rm M}},K_{2{\rm
  M}},K-2,K_{4{\rm M}},K_{5{\rm M}},K_{6{\rm M}})\ , 
  \nonumber 
\end{eqnarray}
and so on. Clearly $\Delta_6(K)\equiv\bar\Delta_6(K)$. 
The differences between $\Delta_i(K)$ and $\bar\Delta_i(K)$ for
$i=2\div 5$ have been found to be negligible. Only 
the differences between $\Delta_1(K)$ and $\bar\Delta_1(K)$ are
sizable. In any case the behavior of
$\bar\Delta_i(K)$ for large $K$ is the same as that discussed for
$\Delta(K)$. Therefore, we propose to estimate the  ``total
missing'' BE by using the formula
\begin{equation}
  (\Delta B)_T= \sum_{i=1,6} c(K_{i{\rm M}},p) \bar \Delta_i(K_{i{\rm
  M}})\ ,
  \label{eq:dbet}
\end{equation}
where $p=5$ ($7$) for the MT-V potential (AV18 and AV18+UIX). To give
an example, the values for $\Delta_i(K_{i{\rm  M}})$ and $c(K_{i{\rm
M}},p)$ computed for the MT-V case are reported in
Table~\ref{tb:estra}, from which it is possible to derive that
$(\Delta B)_T\approx 11$ keV. If this value is added to
$B(72,40,34,28,0,0)=31.347$ MeV, we obtain $31.358$ MeV, which is in
very good agreement with the results obtained by other groups.
For the AV18 potential, Eq.~(\ref{eq:dbet}) gives $(\Delta
B)_T=12$ keV and if this value is added to
$B(72,40,34,28,24,16)=24.210$ MeV, we obtain $24.222$ in close
agreement with the FY estimates of $24.25$ MeV of Ref.~\cite{Nea02}
and $24.223$ of Ref.~\cite{LC04}. Note that
for the class C2 we have computed the coefficient $c(\overline{K},p)$
with $p=7$. Using $p=6$, we have $c(40,6)=3.52$ and $(\Delta B)_2=3.20$ keV 
(instead of $2.60$ keV), a very small change.
For the AV18+UIX model, the same
procedure allows for an extrapolated  BE estimate of $28.474$ MeV, 
again in agreement with the FY value $28.50$ MeV. Note that the FY 
BE results are quoted with an uncertainty of 50 keV due to the 
truncated model space in their calculations~\cite{Nea02}.

\subsection{Results}
\label{sec:resv}

The values obtained for a number of different potential models
after including the states of the 6 different classes up to the
values $K_1=72$, $K_2=40$, $K_3=34$, $K_4=28$, $K_5=24$ and
$K_6=16$ (the last two values only for the realistic cases, for
central potential we have taken $K_5=K_6=0$) are
presented in Tables~\ref{tb:res} and~\ref{tb:resB}. 
Table~\ref{tb:res} presents the 
results for some central potential models, while Table~\ref{tb:resB} 
reports the results for various realistic potentials
with and without including different models for the 3N forces, too. 
The BE's obtained by using the extrapolation technique described
in the previous section are enclosed in parentheses.
Results obtained by other techniques are also reported.

Let us consider first the central potentials (for all of them  we
have taken $\hbar^2/m=41.47$ MeV fm$^2$). We have selected 5
different potential models, i.e. the Volkov~\cite{V65}, the
Afnan-Tang S3 (ATS3)~\cite{AT68}, the Minnesota~\cite{TLT77}, the
MT-V and the Malfliet-Tjon version I/III (MT-I/III)~\cite{MT69}. These
potentials have been 
used by several groups to produce benchmark calculations, but
unfortunately for some of them different versions exist. The
parameters of the first 4 potentials mentioned above can be found in Table~I
of Ref.~\cite{VS95}, while the version of the MT-I/III used has
the same parameters as reported in Table~I of
Ref.~\cite{SSK92}. The Volkov and MT-V are spin-independent,
while the other 3 potentials are spin dependent. Note that it is
customary to include the point-Coulomb interaction ($e^2=1.44$ MeV
fm) with the Minnesota potential, while the MT-I/III version acts
only on s-waves. Clearly for this group of potentials, the total
orbital angular momentum is a good quantum number and therefore we
have included in the WF's only the channels with $L=0$.

The first example is the Volkov potential with Majorana parameter
$M=0.6$.  As can be seen in Table~\ref{tb:res}, our result agrees
very well with the estimates by other techniques, especially with
the one using the SVM~\cite{VS95}. The Volkov potential, given
as a sum of gaussians,  has a very soft core and therefore the
induced two-body correlations in the ground state WF
are weaker than in the other cases. In fact, we have found that
the convergence of the HH expansion is in this case much faster
(it is reached for $K_1\approx 30$).  Since inclusion of HH states
with fairly low  values of the grand angular quantum number are
sufficient to obtain convergence, a successful HH calculation 
for this potential was already possible 20 years ago~\cite{B81} 

Others central potentials often used in the literature are the
ATS3 and Minnesota potentials. Both are given as a sum of gaussians
but have a rather strong repulsion at short interparticle
distances. This induces important two-body correlations in the
WF's and consequently an acceptable convergence for the
first class is reached only for $K_1>40$.  The
chosen version  of the Minnesota potential  has the exchange parameter
$u=1$. As mentioned before, the point-Coulomb potential is
included in the calculation, however, in the WF we have
included only states with $T=0$. In both cases, we observe a good
agreement between the different theoretical estimates.

The three potentials examined so far are given as functions of
gaussians and thus depend on $r_{ij}^2$. As is well known, in such a
case the convergence of the HH expansion as a function of
the grand angular quantum number is exponential and fast. We 
actually observe such a  behavior in all three cases. However, especially
for the class C1, the convergence is relatively more difficult for
the two models with a repulsive core than in the Volkov case,
confirming that it is this class which is mostly responsible for the
need of the two-body correlations.

The next examples considered are the MT-V and MT-I/III potentials.
They are given as a superposition of Yukawians and have a strong
repulsive core with a $1/r$ divergence. As already mentioned they
represent the most challenging  problem for the HH expansion, due
to the difficulty of constructing accurate two-body correlations at
short interparticle distances, where the cancellation between
kinetic and potential energy is critical. As can be seen by
inspecting Table~\ref{tb:res}, the BE for the MT-V is slightly
underestimated. We have already discussed this case in the previous
subsection and we have seen that it is possible to obtain very precise
estimates for the ``missing'' BE using the known behavior
$\Delta\propto 1/{\overline K}^5$.  Adding
this ``missing'' BE  to the value $B=31.347$ MeV brings the HH
results  very close to the estimates computed by other techniques. 
For the (s-wave) MT-I/III we observe that
our estimate is already close to the very precise calculation of
Ref.~\cite{SSK92}. The ``missing'' BE in this case is estimated
to be $21$ keV, bringing our estimated BE to be $30.331$ MeV.

Let us now consider the calculations performed using the realistic
models of the NN interactions (see Table~\ref{tb:resB}). Again
the value $\hbar^2/m=41.47108$ MeV fm$^2$, corresponding to
$2/m=1/m_p+1/m_n$, has been used. Let us consider first the calculations
performed without any 3N interaction. We have considered here the
AV18 and the Nijmegen II \cite{Nijm} (Nijm-II) interactions
models. Both potentials 
belong to the group of the modern NN potentials which reproduce
the NN Nijmegen data set \cite{Nijdata} with a $\chi^2$ per datum
$\approx 1$. 
They have been selected since they are local in
coordinate space, while  other modern potentials either have a
``non-local" term like $\nabla^2$ (Nijmegen I  potential
\cite{Nijm}), or are given in momentum space (Bonn interaction
\cite{Bonn}). Note that our technique does not, in principle, 
present any difficulties in treating these other kind of potentials. The only
problem is that now it is not possible to solve the hyperradial
second order differential equations by the method proposed in Ref.
\cite{KMRV96}. Work is in progress to overcome this difficulty and
to compute the $A=3$ and $4$ WF's also with non-local
potentials in coordinate or momentum space.

The convergence of the HH expansion in the case of the AV18
potential has been already discussed in the previous subsection.
An analogous pattern of convergence is also found for the Nijm-II
potential. In Table~\ref{tb:resB}, the results for the BE
and other properties are compared with the results of other
techniques. Note that in the Nijm-II model we have included 
also the electro-magnetic interactions, in addition to the
Coulomb potential, as in the case of the AV18 potential.
These terms contribute an additional $-0.07$ MeV to the BE
and this explains the difference with the reported FY calculation,
where they were not included.
By taking into account this fact, our Nijm-II BE  agrees well with the
corresponding value obtained using the FY equations. Moreover, by
taking into account the ``missing'' BE estimated as explained
previously,  our results practically reproduce 
the FY ones, by again taking into account
the quoted 50 keV uncertainty of the latter method~\cite{Nea02}.

We now consider the inclusion of the 3N interaction. We have
considered here two models: the already discussed UIX and
Tucson-Melbourne~\cite{TM} (TM) model. In the latter case, we
have used the modified version TM', more 
consistent with chiral symmetry~\cite{FHK99},
with the cutoff parameter fixed
to be $\Lambda=4.756\ m_\pi$~\cite{NKG00}. We have used them
together with the AV18 potential (AV18+UIX and AV18+TM' models).
The cutoff parameter of the TM' 3N interaction was chosen to
reproduce the BE of $\het$. The inclusion of the UIX or the TM'
models of the 3N interaction does not change the convergence behavior
of the HH expansion and also in these cases it is possible to
obtain nearly converged results (they have been obtained in 
the approximation described in Subsec.~\ref{sec:trun}). 
Note that in the AV18+UIX case,
the HH and FY estimates for the BE are slightly above the GFMC
result. This is probably due to fact that in the GFMC technique,
the $L^2$ and $(L\cdot S)^2$ terms of the NN interaction are
not treated exactly and therefore the GFMC estimates have to be
regarded as an upper bound of the true ground state energy.

In summary, 
the HH expansion has proved to be flexible enough to describe
accurately the $\alpha$-particle bound state using realistic NN
and 3N interaction models.

\subsection{Origin of the $T>0$ components}
\label{sec:cib}

In the calculations performed with the realistic NN and NN+3N
interaction we have included components with total isospin
$T=0$, $1$ and $2$ in the WF. The calculated percentages of the
waves with $T=1$ and $T=2$ for the AV18 and AV18+UIX models are
reported in Table~\ref{tb:isom}. The results obtained by the FY
calculations~\cite{Nea02} have been also reported.
These components appear in the
WF when the class C6 is included in the
expansion. From Table~\ref{tb:conv}, it can be seen that the
convergence of the BE for that class is reached without difficulties
including states up to $K=16$. However, the percentage values of
the $T=1$ and $2$ states  have been found to converge substantially more
slowly and HH functions of class C6 up to $K=32$ have to be
considered. The contribution to the BE of the C6 states with $K>16$ is
very small, less than 1 keV.

As can be seen by inspecting Table~\ref{tb:isom}, the percentages of
the components with $T=1$ and $T=2$ in the $\alpha$-particle wave
function are extremely small. For the AV18 potential, 
they are in good agreement with the FY estimates~\cite{Nea02}.
The percentages obtained using the Nijm-II
potential differ by about 40\% with respect to those obtained with
the AV18.
The inclusion of the 3N interaction tends to reduce them slightly.
The adopted models of 3N
interaction do not contain any isospin mixing term.

The knowledge   of the $T=1$ and $2$ percentages is important for
parity violating experiments of electron scattering on $\heq$, devoted
to studying the admixture of strange quark $s\bar s$ pairs
in the nucleon. Information on this quantity can be extracted from
the measurement of the ``left-right'' asymmetry $A_{LR}$ of polarized
electrons on a target nucleus, resulting from the interference
between the electromagnetic and the weak neutral current
mediating the scattering process. The study of the asymmetry is particularly
simple in case of a $(J^\pi,T)=(0^+,0)$ system, since in that
case the number of matrix elements entering this observable is small.
Moreover, the use of $\heq$ as a target nucleus is also favored
by the fact that its first excited state is at $20.1$ MeV, 
which allows for an easy experimental control of
inelastic processes. Indeed, there are approved experiments
at the Jefferson Lab~\cite{BmK01,Happex}.

However, the extraction of the information from the experiments 
could be complicated by the presence of components with 
isospin $T=1$ and $2$in the WF of $\heq$.
This question was analyzed in Ref.~\cite{RHD94} and found that 
the contribution from the $T=1$ isospin mixing configurations to $A_{LR}$
was negligible. Considering the effect of the Coulomb
potential alone, the percentage of the $T=1$ component 
in that work was estimated to be $P_{T=1}=7\times 10^{-4}$.
From the present calculation, in agreement with 
the study of Ref.~\cite{Nea02}, the $T=1$
component results to be is 4 times larger and this could be 
of some effect on $A_{LR}$. 

It is interesting to study the origin of the $T=1$ and
$2$ isospin admixtures to the $\alpha$ particle WF. To this end
we have  performed a series of calculations by removing 
from the Hamiltonian the different terms which induce the $T>0$
components. Let us write 
\begin{equation}
  H = H_{\rm IC} + H_C + H_{\rm CSB} + H_{\rm e.m.} + K_{\Delta}
  \ ,\label{eq:isom}
\end{equation}
where $H_{IC}$ is the isospin-conserving part of the nuclear
Hamiltonian, $H_C$ the point-Coulomb interaction, $H_{\rm CSB}$
the charge symmetry breaking nuclear interaction  (namely, the
operators 15-18 in AV18), $H_{\rm e.m.}$ the remaining
electro-magnetic (e.m.) interaction (finite-size effects, vacuum
polarization, magnetic moment 
interactions, etc) and $K_{\Delta}$  the term originating from
the  proton and neutron mass difference in the kinetic energy.
This latter term has not been included in the solution of the
four body problem and its effects have been evaluated perturbatively
as explained below.

By approximating the Hamiltonian with $H_{\rm IC}$ only,
one would get no isospin admixture at all. We have then
added the various terms one by one to $H_{\rm IC}$, and reported
the results in Table~\ref{tb:isom2}. As can be seen from that
table, with the inclusion of the Coulomb potential $H_C$,
the percentage of the $T=1$ state is in rough agreement (within a
factor 2) with that estimated in  Ref.~\cite{RHD94}.  
The percentage of the $T=2$ state is very tiny
in this case. When the CSB terms of the AV18 NN interaction are taken 
into account, however, the previous picture is noticeably modified and
both components increase, in particular the 
$T=2$ component which becomes larger than the $T=1$ one. 
Finally, the effect of $H_{\rm e.m.}$ is rather tiny.

To further analyze 
the origin of the isospin admixture components, we  have repeated the
approximate calculation of Ref. \cite{RHD94}.  Let us write
\begin{equation}
  H_C + H_{\rm CSB} + H_{\rm e.m.} + K_{\Delta} 
     \equiv H_{I}^{(0)}+H_{I}^{(1)}+
   H_{I}^{(2)}\ ,   \label{eq:cib}
\end{equation}
namely as a sum of an isoscalar, isovector and isotensor
term (the isoscalar part comes from the Coulomb and e.m. potentials).
Let us treat  $H_{IC}+H_{I}^{(0)}$ as the unperturbed Hamiltonian,
and try to evaluate the $T>0$ components using first order
perturbation theory. Namely,
\begin{equation}
  |\delta\Psi^{(T)}\rangle = \sum_{n>0} |\Psi_n \rangle\;
  {\langle \Psi_n |H_{I}^{(T)}| \Psi_0 \rangle \over E_0-E_n}\ ,
  \label{eq:fior}
\end{equation}
where $\Psi_0$ is the unperturbed ground state and $\Psi_n$,
$n=1,2,\ldots$ the unperturbed excited states of
$H_{IC}+H_{I}^{(0)}$, which therefore have definite values of the
total isospin quantum number. In particular,  $\Psi_0$ has $T=0$,
etc. The most important contributions to the components
$T=1$, $2$ of $\delta\Psi^{(T)}$ would come from the
lowest excited states. Following Ref. \cite{RHD94} (see also
Ref.~\cite{BM69}), we model these 
states as
\begin{equation}
   |\Psi_1^{(T)} \rangle=  {\Omega_T| \Psi_0\rangle
          \over 
      \langle \Psi_0 | \Omega_T^\dag  \Omega_T| \Psi_0\rangle^{1/2}  }
   \ ,  \label{eq:fior2}
\end{equation}
where $\Omega_T$, $T=1,2$, are excitation operators of the form
\begin{eqnarray}
  \Omega_1&=&\sum_{ij} r^2_{ij} \Bigl (\tau_z(i)+\tau_z(j)\Bigr)\ ,
  \label{eq:fior3}\\
  \Omega_2&=&\sum_{ij} r^2_{ij} \Bigl ( \tau_z(i)\tau_z(j) -(1/3)
  \bmta(i)\cdot\bmta(j)\Bigr)\ .
  \label{eq:fior4}
\end{eqnarray}
The operator $\Omega_1$ generates a state $(J^\pi,T)=(0^+,1)$
corresponding to a ``breathing'' mode where neutrons and
protons oscillate in counter-phase. Furthermore, $\Omega_2$ generates
a state $(J^\pi,T)=(0^+,2)$ with ``tensor'' oscillations. The $T=1,2$
components in the $\heq$ wave 
function would be given by
\begin{equation}
  |\delta\Psi^{(T)}\rangle \approx  |\Psi_1^{(T)} \rangle\;
  {\langle \Psi_1^{(T)} |H_{I}^{(T)}| \Psi_0 \rangle \over E_0-E_1(T)}
   \equiv \chi_T |\Psi_1^{(T)} \rangle\ ,
  \label{eq:fior5}
\end{equation}
where $E_1^{(T)}=\langle \Psi_1^{(T)} |H_{IC}+H_{I}^{(0)}| \Psi_1^{(T)}
\rangle$. The percentage of the $T$-wave is just
$100|\chi_T|^2$. 

In the following, we applied the procedure outlined above using the
AV18 potential model. $\Psi_0$ is the WF computed with the
HH expansion excluding any states belonging to the class C6. We have found
\begin{equation}
  E_0=-24.19\ {\rm MeV}\ , \qquad
  E_1^{(1)}=7.05\ {\rm MeV}\ ,\qquad 
  E_1^{(2)}=29.07\ {\rm MeV} \ .
  \label{eq:fior6}
\end{equation}

Let us first consider the (point) Coulomb potential $H_C$, given by 
\begin{equation}
  H_C=\sum_{i<j} {e^2\over r_{ij}}
  \left( { 1+\tau_z(i)\over 2} \right ) \,
  \left( { 1+\tau_z(j)\over 2} \right ) \,
  \label{eq:coou}
\end{equation}
and therefore
\begin{eqnarray}
  H_{I,C}^{(1)}&=&\sum_{i<j} {e^2\over r_{ij}}
  \left( { \tau_z(i)+\tau_z(j)\over 4} \right ) \ ,
  \label{eq:coou2} \\
  H_{I,C}^{(2)}&=&\sum_{i<j} {e^2\over r_{ij}}
  \left( { \tau_z(i)\tau_z(j)-{\scriptstyle 1\over \scriptstyle3 } 
   \bmta(i)\cdot\bmta(j)\over
  4} \right ) \ .   \label{eq:coou3}
\end{eqnarray}
The necessary matrix elements can be readily computed with the result 
that
\begin{equation}
  \langle \Psi_1^{(T=1)} |H_{I,C}^{(T=1)}| \Psi_0 \rangle\approx -100\
  {\rm keV}  \ ,\qquad
  \langle \Psi_1^{(T=2)} |H_{I,C}^{(T=2)}| \Psi_0 \rangle\approx  -37\ 
   {\rm keV} \ .
  \label{eq:coou4}
\end{equation}
and therefore $P_{T=1}\approx 1\times 10^{-3}$  and 
$P_{T=2}\approx 0.05\times 10^{-3}$ confirming that the $T=2$ component
induced by the Coulomb potential is smaller than the $T=1$ one.
In fact, the radial dependence (and strength) of  $H_{I,C}^{(1)}$ and
$H_{I,C}^{(2)}$ are the same. However, the $T=2$ breathing
mode has a higher excited energy and this reduces the probability
of a  ``transition'' to the state $|\Psi_1^{(T=2)}\rangle$.
The values $P_{T=1}$ and $P_{T=2}$ are also in rough
agreement (within a factor 2) with the results reported in the second
row of Table~\ref{tb:isom2}. It appears that in our approximate calculation,
the $T=1$ and $2$ percentages are somewhat  underestimated. Note that
the value $P_{T=1}\approx 1 \times 10^{-3}$ is also in reasonable
agreement with the  estimate of Ref.~\cite{RHD94}.

As seen before, this situation is reversed when the CSB part of the
nuclear interaction is taken into account. 
The CSB part of the AV18 potential is explicitly given by
\begin{equation}
  H_{CSB} =\sum_{i<j}
  \Bigl( V^{(1)}(i,j)   [\tau_z(i)+\tau_z(j)] +  V^{(2)}(i,j)
  [\tau_z(i)\tau_z(j)-{\scriptstyle 1\over\scriptstyle
  3}\bmta(i)\cdot\bmta(j)]\Bigr)   \equiv 
  H_{I,CSB}^{(1)}+  H_{I,CSB}^{(2)}\ ,
  \label{eq:csbb}
\end{equation}
where $ V^{(1)}(i,j)$ and $ V^{(2)}(i,j)$ are functions depending
on the interparticle distance $r_{ij}$ and on the spin operators
of the particles $i$ and $j$. The operator $
[\tau_z(i)\tau_z(j)-{1\over 3}\bmta_i\cdot\bmta_j] $ in
$H_{I,CSB}^{(2)}$ induces differences between the $pp$  and $pn$
interactions (it originates mainly from the difference between the
charged and neutral pion masses). These differences are well
established and, although rather small, are of sizable value in
some observables (such as the singlet $np$ and $pp$ scattering lengths).
The operator $ [\tau_z(i)+\tau_z(j)]$ in
$H_{I,CSB}^{(1)}$ induces instead differences between the $pp$ and $nn$
interaction, too. Due to the lack of precise $nn$ data, the magnitude
of this charge independence breaking term is not very well known,
however, its strength  should satisfy $H_{CSB}^{(1)}\ll H_{CSB}^{(2)}$. 
Turning to our approximate method, we have found that
\begin{equation}
  \langle \Psi_1^{(T=1)} |H_{I,CSB}^{(T=1)}| \Psi_0 \rangle\approx -23
  \ {\rm keV}\ ,\qquad
  \langle \Psi_1^{(T=2)} |H_{I,CSB}^{(T=2)}| \Psi_0 \rangle\approx -84
  \ {\rm keV}\ .\label{eq:csbb2}
\end{equation}
As expected, the matrix elements $\langle \Psi_1^{(T=2)} |H_{I,CSB}
^{(2)}| \Psi_0 \rangle$ are larger than the corresponding 
matrix elements of $H_{I,CSB}^{(1)}$. Adding the values given in
Eqs.~(\ref{eq:coou4}) and~(\ref{eq:csbb2}), we find $P_{T=1}=1.5\times
10^{-3}$ and $P_{T=2}=0.6\times 10^{-3}$ as the percentage of the isospin
components induced by the Coulomb+CSB part of the Hamiltonian. 
Our approximate calculation seems to underestimate both
component percentages (in particular $P_{T=2}$), but it is in
qualitative agreement with the results reported in Table~\ref{tb:isom2}.

The e.m. part of the Hamiltonian produces small effects. This is also
supported by our approximate calculation. We have found, in fact, that
\begin{equation}
  \langle \Psi_1^{(T=1)} |H_{\rm e.m.}| \Psi_0 \rangle\approx 6
  \ {\rm keV}\ ,\qquad
  \langle \Psi_1^{(T=2)} |H_{\rm e.m.}| \Psi_0 \rangle\approx -3
  \ {\rm keV}\ .\label{eq:cvem}
\end{equation}

The last effect which should be taken into consideration is the
mass difference between neutrons and protons. The associated term
in the Hamiltonian is
\begin{equation}
  H_{\Delta } =\sum_{i=1,4}
  {1\over 2} \left( {{1\over 2m_p}- {1\over 2 m_n} }\right )
  \nabla_i^2 \;\; \tau_z(i) \equiv H_{I,\Delta}^{(1)}\ ,
  \label{eq:dima}
\end{equation}
and therefore $H_{\Delta}$ is an isovector operator and may
increase the percentage of the $T=1$ state.
We have not considered this term in the solution of the full
four-body problem but we have estimated its contribution
using the approximated procedure outlined before. We have found that
\begin{equation}
  \langle \Psi_1^{(T=1)} |H_{I,\Delta}^{(T=1)}| \Psi_0 \rangle\approx
  -9\ {\rm keV}\ .
  \label{eq:dima2}
\end{equation}
The effect of $H_{\Delta}$ is therefore rather small (it would produce
a change of $P_{T=1}$ of about $0.1\times 10^{-4}$).

This study therefore roughly support the result of our full
calculation, namely that the charge symmetry breaking
terms of the nuclear interaction are responsible for the
relatively large $T=2$ components in the $\alpha$-particle wave
function. On the contrary, the $T=1$ component originates mainly
from the Coulomb interaction. The other e.m. terms and the
neutron-proton mass difference play a negligible role.

\subsection{Truncation studies}
\label{sec:trun}

In prevision of future applications of the present technique to
heavier systems, we have explored the effect of truncating part of
the NN and 3N interaction. The aim is to find a way of simplifying
the Hamiltonian, still obtaining very precise results, with a
maximum deviation of the order of 
0.1\% with respect to the results obtained without any
approximation. We have explored both the effect of neglecting the NN
interaction when the total angular momentum  $j$ of the pair is
greater than a given value $j_{\rm M}$ and the effect of
neglecting the 3N interaction when it acts on HH states with grand
angular quantum number greater than a given $K_{\rm M}$.

Let us first discuss the case of truncation of the NN interaction. 
We have considered an NN interaction which vanishes when acting on 
pairs with total angular momentum $j>j_{\rm M}$, and have varied $j_{\rm
M}$ to see the effect on the $\alpha$-particle BE. We have
considered the AV18 interaction, since its operatorial form allows
one to compute it for arbitrary values of $j$. The results obtained
can be seen in Table \ref{tb:truj}. The calculation with $j_{\rm
M}=\infty$ means that we have retained the NN interaction acting
in all states. By inspecting the Table, it can be seen that
by taking $j_{\rm M}\ge 6$ the BE and other quantities vary very little.
Therefore, it seems safe to retain the NN interaction as acting only on
states with $j\le 6\div 8$.

Let us now consider the 3N interaction. The behavior of
the radial parts of the UIX or TM' 3N potential
are rather soft  at short interparticle distances. Since the
large grand angular quantum number components  in the WF
are induced by the repulsive core of the potentials, this
suggests that the correlations induced by the 3N interaction
would not need such high components. Therefore, we have included
in the Hamiltonian an effective 3N interaction of the kind
\begin{equation}
  \widetilde W_{K_{\rm M}}(i,j,k) =
  P_{K_{\rm M}}^\dag W(i,j,k)  P_{K_{\rm M}}
  \label{eq:truk}
\end{equation}
where $W(i,j,k)$ is one of the standard 3N interactions and
$P_{K_{\rm M}}$ a projection operator which gives 0 when it acts on 
four body HH states with a given grand angular quantum number $K$  if
$K>K_{\rm M}$. Actually, $\widetilde W_{K_{\rm M}}(i,j,k)$ is an effective 
4-body interaction. We have then studied the effects on the
$\alpha$-particle BE by varying $K_{\rm M}$. We have considered
here the AV18+UIX model. The results obtained can be found in
Table~\ref{tb:truk}, where for simplicity we have restricted the HH
basis to include only the first three classes (in any case, the other
classes include HH states with $K<30$). As can be seen by inspecting
the Table, the BE and the other quantities depend very little
on $K_{\rm M}$. Already for $K_{\rm M}=20$, the corresponding
BE differs from that obtained in the non-truncated case by
less than 0.1\%. The calculation of the 3N potential matrix
elements between states with $K\le 20\div 30$ is noticeably simpler than
in the general case and still the results are of acceptable
precision. This should allow for HH calculations including 3N
interactions also for scattering states and for heavier systems.
This conclusion is also supported by the study of Ref.~\cite{Bea04}
of the incorporation of the 3N interaction in the EIHH method.
Note that the results reported in Tables~\ref{tb:conv} and~\ref{tb:resB}
have been obtained using the 3N interactions ``truncated''
as in Eq.~(\ref{eq:truk}) and taking $K_M=30$.

\subsection{Asymptotic Normalization Constants}
\label{sec:anc}

The asymptotic normalization constants (ANC's) are properties
of the bound state WF which can be related to experimental
observables. They are interesting quantities from
which useful information on the nuclear structure can be 
extracted.
They also provide a test of the
quality of the variational WF in the asymptotic region, as we shall see.
This test will be particularly severe in our approach,
as the description of the $\heq$ WF  in terms of the four-body HH
functions in regions where the 1+3 or 2+2 clustering configurations 
are dominant will be difficult.

Let us concentrate first on
the proton-triton ANC $C_S^{pt}$ of $\heq$, defined by
\begin{equation}
  \Psi_4(\jacb_{1},\jacb_{2},\jacb_{3})
   \rightarrow C_S^{pt} \sqrt{2\beta_{pt}}\; 
   {W_{-\eta_{pt},1/2}(2\beta_{pt} r_{pt})
  \over r_{pt} }\; \Phi^{pt}_0(\widehat\jacb_{1},\jacb_{2},\jacb_{3})\ ,
  \qquad r_{pt}\rightarrow\infty\ ,  \label{eq:anc1}
\end{equation}
where the Jacobi vectors $\jacb_i$ correspond to the permutation $p=1$ 
(the index $p$ will be suppressed in this section)
and $r_{pt}=\sqrt{2/3}\; \jacb_{1}$ is the distance between the
$\tri$ center of mass and the fourth nucleon.
The function $\Phi^{pt}_L$ is defined as
\begin{equation}
  \Phi^{pt}_L(\widehat\jacb_{1},\jacb_{2},\jacb_{3})=
   \biggl\{ Y_L(\widehat\jacb_{1}) \Bigl[ \psi_t(1,2,3)\chi_4\xi_4
   \Bigr]_S\biggr\}_{0,0} \ , \label{eq:anc2}
\end{equation}
where $\psi_t(1,2,3)\equiv\psi_t(\jacb_{2},\jacb_{3})$ is the
$\tri$ WF and $\chi_4$ ($\xi_4$) is the spin (isospin) function of the
fourth nucleon. In the previous equation, the spin $1/2$ of $\tri$ is coupled 
to the spin $1/2$ of the other nucleon to give a ``channel'' spin $S=0,1$. 
The channel spin is in turn coupled to $L$ to give a total angular momentum
$J=0$, therefore $L=S$. Due to the even parity of the $\heq$ state,
the $p-\tri$ clusters can only be in the state $S=L=0$.  
In Eq.~(\ref{eq:anc1}), $W_{-\eta,j}(2\beta r)$ is the Whittaker function
which behaves irregularly at the origin and decays exponentially for 
$r\rightarrow\infty$, while $\beta_{pt}$ and $\eta_{pt}$ are determined by
\begin{equation}
   \beta_{pt}=\sqrt { {3\over 2} {m\over\hbar^2} (B_4-B_t)}\ ,\qquad
   \eta_{pt}={3\over 4} {m\over \hbar^2} {e^2\over \beta_{pt}}\ ,
    \label{eq:anc3}
\end{equation}
where $e^2\approx1.44$ MeV fm,  $\hbar^2/m\approx 41.47$ MeV fm$^2$
and $B_4$ and $B_t$ are the $\heq$ and $\tri$ BE,
respectively.
Finally, the factor $\sqrt{2\beta_{pt}}$ in Eq.~(\ref{eq:anc1}) 
has been introduced so that the ANC $C_S^{pt}$ be adimensional.

\begin{figure}[p]
\includegraphics[scale=.9]{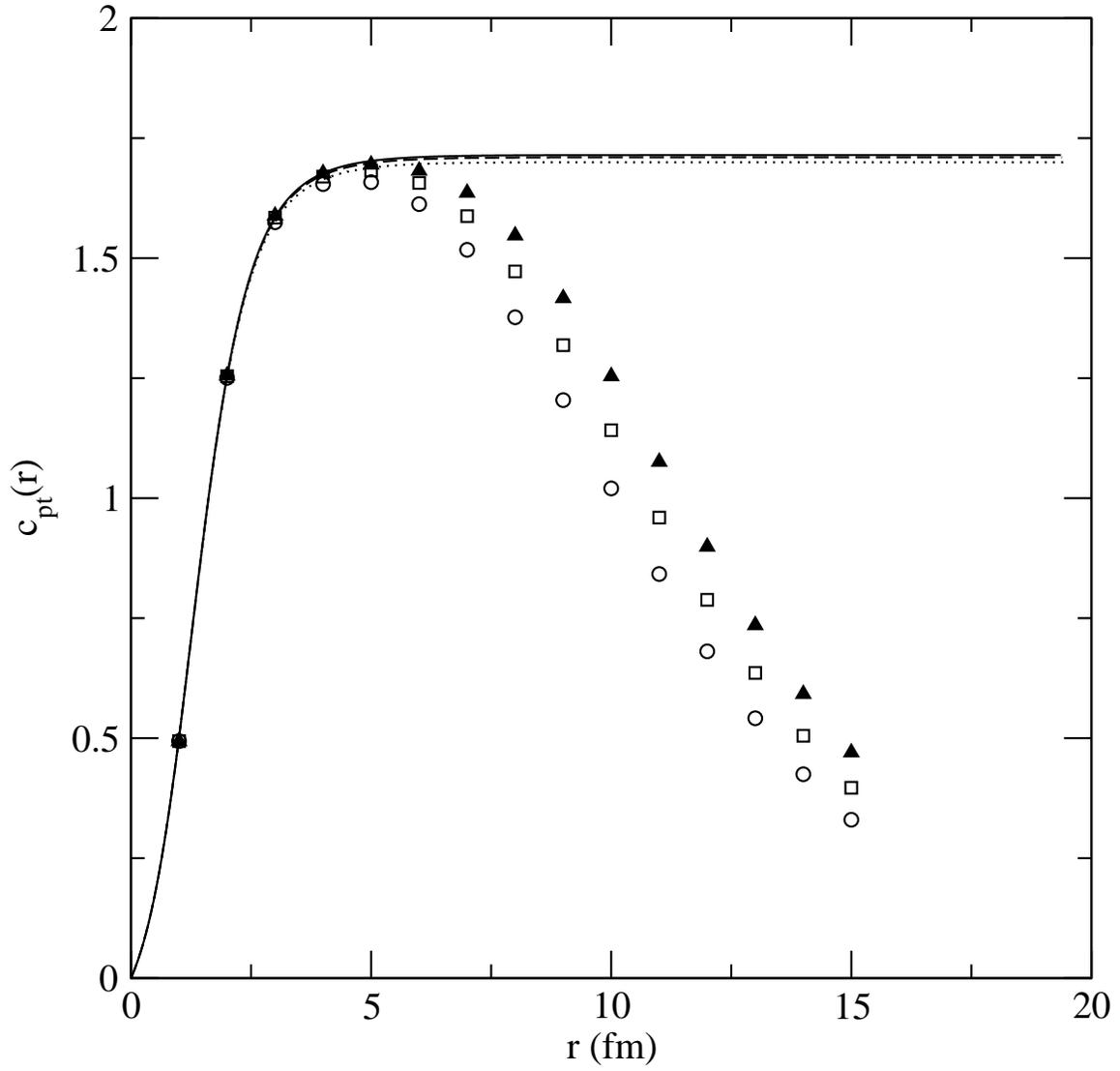}
\caption{Ratios $c_{pt}(r_{pt})$ as function of the
$p-\tri$ distance $r_{pt}$. The ratios
obtained by the direct calculation of the overlap
defined in Eq.~(\ref{eq:anc4}) with the $\heq$ WF 
corresponding to the gran angular quantum numbers specified 
in Eq.~(\ref{eq:anc66}) are shown by the open circles (case a), 
the open squares (case b) and the solid triangles (case c).
The ratios obtained by the solution of the differential equation
defined in Eq.~(\ref{eq:anc7}) are shown by the dotted (case a), 
dashed (case b) and solid lines (case c), respectively (the dashed 
and solid line are almost coincident). The $\heq$ WF were generated
using the AV18 potential.
}
\label{fig:anc}
\end{figure}

In order to calculate $C_S^{pt}$, let us introduce the $\tri-\heq$ overlap
function as
\begin{equation}
  f_{pt}(r_{pt})=\int d^3\jac_{1}\; d^3\jac_{2}\; d^3\jac_{3}\;
    \delta\Bigl(\sqrt{2\over3} \jac_{1}-r_{pt}\Bigr)
    {  \Phi^{pt}_0(\widehat\jacb_{1},\jacb_{2},\jacb_{3})^\dag\over r_{pt}}
     \Psi_4(\jacb_{1},\jacb_{2},\jacb_{3}) \ ,
    \label{eq:anc4}
\end{equation}
and the ratio
\begin{equation}
   c_{pt}(r_{pt}) = { f_{pt}(r_{pt}) \over \left(3/2\right)^{3\over 2}
     \sqrt{2\beta_{pt}}\; W_{-\eta_{pt},1/2}(2\beta_{pt} r_{pt})}\ .
    \label{eq:anc5}
\end{equation}
If $\Psi_4$ is the exact $\heq$ WF, for $r_{pt}\rightarrow\infty$ 
the overlap function behaves as
\begin{equation}
   f_{pt}(r_{pt})\rightarrow \left({3\over 2}\right)^{3\over 2}
     C_S^{pt}  \sqrt{2\beta_{pt}}\;
     W_{-\eta_{pt},1/2}(2\beta_{pt} r_{pt})\ ,
    \label{eq:anc6}
\end{equation}
and therefore $c_{pt}(r_{pt})\rightarrow C_S^{pt}$,
allowing for the extraction of the ANC. The $\tri$
WF has been determined by means of the pair-correlated HH (PHH)
technique described in Ref.~\cite{KRV94} and is believed to be very 
precise~\cite{Nea03}. The dependence
to the truncation level of the HH basis used to compute 
the $\heq$ WF has been studied by computing the overlap function
for the following three different choices of 
the maximum values of the grand angular quantum numbers of the six
classes as defined in Subsec.~\ref{sec:conv},
\begin{eqnarray}
  \{ K_{1M},K_{2M},K_{3M},K_{4M},K_{5M}, K_{6M}\}&=&
        \{56,32,26,16,14,24\}\ , \qquad {\rm case\ a}\ , \nonumber \\
  \{ K_{1M},K_{2M},K_{3M},K_{4M},K_{5M}, K_{6M}\}&=&
        \{60,36,30,20,18,28\} \ , \qquad {\rm case\ b}\ , \label{eq:anc66} \\
  \{ K_{1M},K_{2M},K_{3M},K_{4M},K_{5M}, K_{6M}\}&=&
        \{64,40,34,24,22,32\}\ , \qquad {\rm case\ c}\ . \nonumber
\end{eqnarray}
The corresponding ratios $c_{pt}$
are shown in  Fig.~\ref{fig:anc} by
the open circles (case a), open squares (case b) and 
solid triangles (case c). The potential used to generate
the WF is the AV18 interaction.
As can be seen by inspection of the figure,
all three functions $c_{pt}$ start to deviate
from the expected asymptotic constant behavior
already for $r_{pt}>5$ fm, showing the 
difficulty of reproducing the cluster structure of the WF 
by means of the four-body HH functions. From the differences between
the three ratios, the very
slow convergence of $c_{pt}(r)$ as a function of
$\{ K_{1M},K_{2M},K_{3M},K_{4M},K_{5M}, K_{6M}\}$
results to be evident. In particular,
a detailed analysis has shown that the convergence is
sensitive to the value of $K_{1M}$. The ratio $c_{pt}(r)$ obtained with 
the larger basis shows a slightly larger
``plateau'' around $r_{pt}=5$ fm, allowing for a crude estimate
of the ANC, $C_S^{pt}\approx 1.7$

\begin{figure}[p]
\includegraphics[scale=.9]{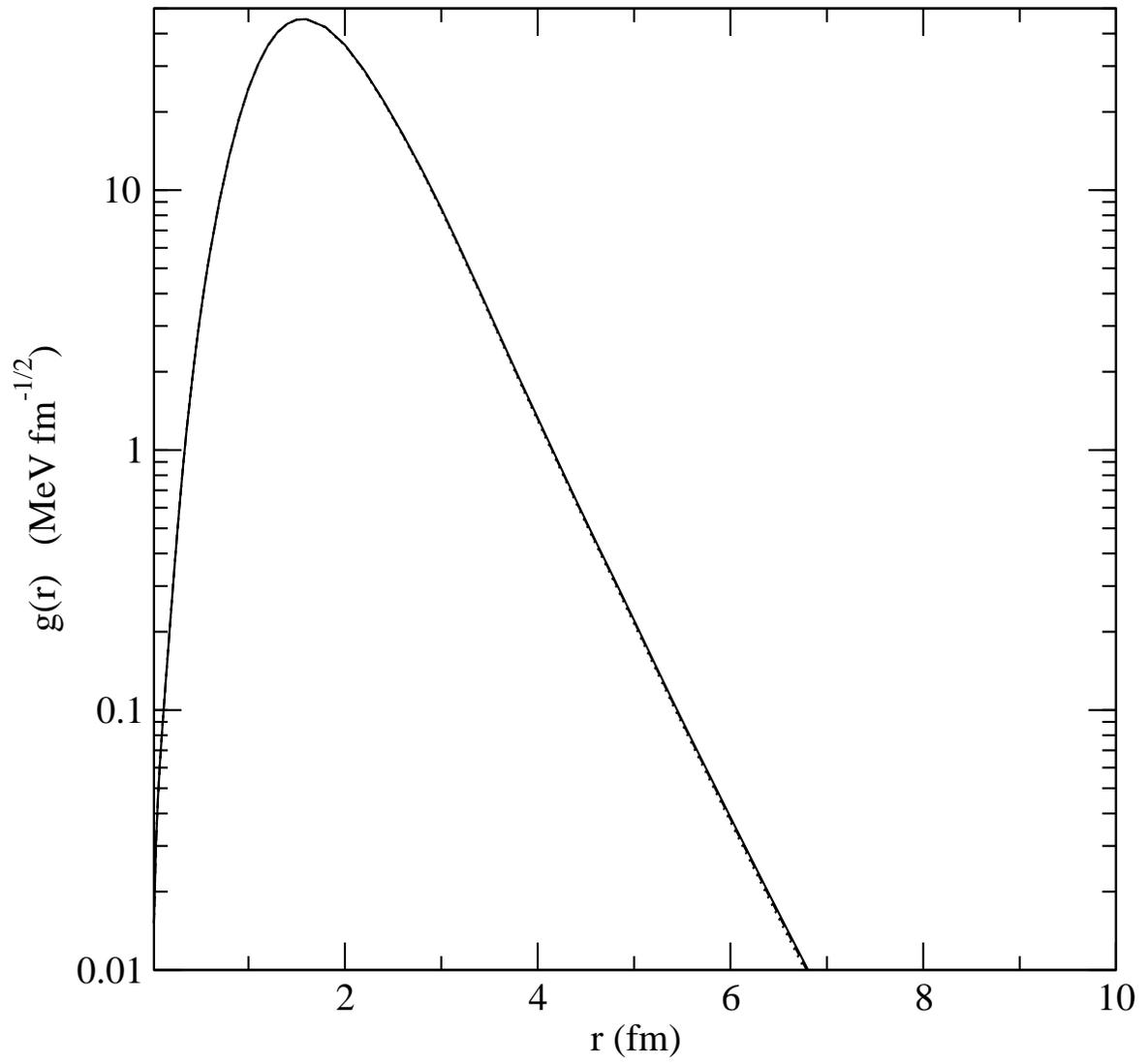}
\caption{Functions $g(r)$ obtained for three choices of the HH basis
specified  in Eq.~(\ref{eq:anc66}) are shown by the dotted (case a), 
dashed (case b) and solid lines (case c).
The three lines are pratically coincident and cannot
be distinguished.
}
\label{fig:g}
\end{figure}

To obtain a greater accuracy in the extraction of the ANC
we have followed another procedure~\cite{T98}. Assuming that $\Psi_4$ 
and $\psi_t$ are ``exact'', it is 
not difficult to show that the overlap function should satisfy
the following differential equation
\begin{equation}
  -{2\over 3}\htm f_{pt}''(r)+{e^2\over r}f_{pt}(r)
   +(B_4-B_t) f_{pt}(r) + g(r) =0 \ ,
  \label{eq:anc7}
\end{equation}
where $r\equiv r_{pt}$ and
\begin{eqnarray}
  g(r)&=&\int d^3\jac_{1}\; d^3\jac_{2}\; d^3\jac_{3}\;
       \delta\Bigl(\sqrt{2\over3} \jac_{1}-r\Bigr)
    {  \Phi^{pt}_0(\widehat\jacb_{1},\jacb_{2},\jacb_{3})^\dag \over r_{pt}}
     \Bigl[ V_{14}+V_{24}+V_{34} \nonumber\\
     && \qquad\qquad +W_{124}+W_{134}+W_{234}-{e^2\over r}
     \Bigr]
     \Psi_4(\jacb_{1},\jacb_{2},\jacb_{3}) \ ,
  \label{eq:anc8}
\end{eqnarray}
$V_{ij}$ and $W_{ijk}$ being the NN and 3N potential, respectively.
As $r\rightarrow\infty$, the function $g(r)\rightarrow 0$,
and the solution of Eq.~(\ref{eq:anc7}) coincides with the Whittaker
function, allowing for the extraction of the ANC via Eq.~(\ref{eq:anc5}).
We have computed the function $g(r)$ with the three different choices 
of the $\heq$ WF $\Psi_4$ given in Eq.~(\ref{eq:anc66}) and reported 
the results in Fig.~\ref{fig:g}. 
As can be seen, the function $g(r)$ is peaked at $r\approx 2$ fm,
goes to zero exponentially and depends slightly on the choice 
of $\Psi_4$. In fact,  the selected HH bases are already
large enough to accurately describe the $p-\tri$ decomposition 
for $r< 4 $ fm. For larger distances, probably $g(r)$ is not 
computed accurately using our variational WF, but there $g(r)$ becomes
vanishingly small and the resulting effect on the ANC is negligible.
This has been checked explicitly by solving Eq.~(\ref{eq:anc7})
(imposing the boundary conditions $f_{pt}(0)=f_{pt}(\infty)=0$)
and computing $c_{pt}(r)$ for the same three cases as before. The results
for $c_{pt}(r)$ are shown in Fig.~~\ref{fig:anc} by the dotted, dashed and
solid lines (the results of the latter two cases are practically 
indistinguishable). As expected, for $r<5$ fm, the line
goes through the ratio functions $c_{pt}(r)$ computed directly 
via the overlap integral and reaches a constant value, corresponding 
to $C_S^{pt}$, around $r=5$ fm.
The extraction of the ANC can be now achieved without any difficulty
and the value found is $C^{pt}_S=1.715$.

An analogous procedure have been repeated for the $n-\het$ ANC and the 
S- and D-wave $d-d$ ANC's. We have used the definition
\begin{eqnarray}
  \Psi_4&\rightarrow& C_S^{nh} \;\sqrt{2\beta_{nh}}\;
     {e^{-\beta_{nh} r_{nh}}
  \over r_{nh} }\; \Phi^{nh}_0(\widehat\jacb_1,\jacb_2,\jacb_3)\ ,
  \qquad r_{nh}=\sqrt{2\over 3} \jac_1 \rightarrow\infty\ , \label{eq:new2} \\
  \Psi_4&\rightarrow& C_S^{dd} \;\sqrt{2\beta_{dd}}\;
       {W_{-\eta_{dd},1/2}(2\beta_{dd} r_{dd})
  \over r_{dd} }\; \Phi^{dd}_0(\jacb_1',\widehat\jacb_2',\jacb_3') \\
   &+& C_D^{dd} \;\sqrt{2\beta_{dd}}\;
      {W_{-\eta_{dd},5/2}(2\beta_{dd} r_{dd})
  \over r_{dd} }\; \Phi^{dd}_2(\jacb_1',\widehat\jacb_2',\jacb_3')\ ,
  \qquad r_{dd}=\sqrt{1\over 2}\jac_2'\rightarrow\infty\ ,\label{eq:new3}
\end{eqnarray}
where $\jacb_i'$ are the set B of the Jacobi vectors defined 
in Eq.~(\ref{eq:JcbV2}) corresponding to the permutation $p=1$ and
\begin{eqnarray}
  \Phi^{nh}_L(\widehat\jacb_{1},\jacb_{2},\jacb_{3})&=&
   \biggl\{ Y_L(\widehat\jacb_{1}) \Bigl[ \psi_h(1,2,3)\chi_4\xi_4
   \Bigr]_S\biggr\}_{0,0} \ , \label{eq:new4} \\
  \Phi^{dd}_L(\jacb_{1}',\widehat\jacb_{2}',\jacb_{3}')&=&
   \biggl\{ Y_L(\widehat\jacb_{2}') \Bigl[ \phi_d(1,2)\phi_d(3,4)
   \Bigr]_S\biggr\}_{0,0} \ . \label{eq:new5}
\end{eqnarray}
In the latter, $\psi_h$ and $\phi_d$ are the $\het$ and deuteron WF, 
respectively, and
\begin{equation}
   \beta_{nh}=\sqrt { {3\over 2} {m\over\hbar^2} (B_4-B_h)}\ ,\qquad
   \beta_{dd}=\sqrt { {2m\over\hbar^2} (B_4-2B_d)}\ ,\qquad   
   \eta_{dd}={m\over \hbar^2} {e^2\over \beta_{dd}}\ ,
    \label{eq:new6}
\end{equation}
with $B_h$ and $B_d$ the $\het$ and deuteron BE, respectively.
For the $d-d$ case, one can also estimate the
distorted-wave parameter $D_2$ defined by
\begin{equation}
  D_2={1\over 15} \int_0^\infty dr_{dd}\; r_{dd}^3 \; f_{dd}^D(r_{dd}) 
    / \int_0^\infty dr_{dd}\; r_{dd} \; f_{dd}^S(r_{dd})\ ,
   \label{eq:anc9}
\end{equation}
where $ f_{dd}^X(r_{dd})$ ($X=S$, $D$) are the S- and D-wave 
$(d-d)-\heq$ overlap functions, respectively, defined
in analogy to Eq.~(\ref{eq:anc4}). The results obtained have been
reported in Table~\ref{tb:anc}, together with some other 
theoretical and experimental estimates available for $D_2$
(for a more complete list of references, see Ref.~\cite{adhi94}).
The $D_2^{dd}$ parameter was determined  in Ref.~\cite{adhi94}
using an approximated method (a cluster model) which however seems 
to provide an estimate rather close to ours. This parameter is also
in reasonable agreement with the experimental values reported in 
Table~\ref{tb:anc}, also considering the difficulty of the extraction
of this quantity from the experimental data.

\section{Conclusions and perspectives}
\label{sec:disc}

We have studied the solution of the Schroedinger equation for the 
four-nucleon ground state using the HH function expansion. 
The main difficulty when using the HH basis is its large degeneracy, 
accordingly a suitable selection of
the HH  functions giving the most important contributions has to 
be performed. In this work, the HH functions have been divided into
classes, depending on the number of correlated particles, the values
of the orbital angular momenta, the total isospin quantum number, etc.
For each class, the expansion has been truncated so as to obtain the 
required accuracy. We have applied this procedure in particular to 
the study of the ground state of the $\alpha$-particle using a number of
NN and NN+3N interaction models.
In all the cases, accurate calculations of the
BE and other ground state properties, such as the asymptotic normalization
constants, have been achieved. 

A similar procedure can be also applied for solving scattering problems. 
The calculation of the phase shifts and the various observables 
for $n-\tri$ and $p-\het$ elastic scattering 
is now in progress and will be published elsewhere~\cite{VKR03S}.

The hyperspherical formalism is adequate for treating all kinds of modern
potentials, except those containing a hard-core.
We have considered here the AV18 and Nijmegen-II NN
potentials and the UIX and TM' 3N interactions.
The inclusion of the $\nabla^2$ term present in the
Nijmegen-I~\cite{Nijm}
potential does not introduce additional difficulties.
As an example, such a term was taken into account in ref.~\cite{RVK95}
where the PHH approach was used. Moreover, the HH method can be
easily formulated in momentum space. It can therefore be applied also
to the case of the Bonn potential~\cite{Bonn} although one additional numerical
integration and the solution of an integral equation are then required.
The application of the HH technique to the $A=3$, $4$ systems  with
the Bonn potential is actually under way.

At present there are only a few other methods
available for accurate calculations of the four-nucleon problem, in
particular by taking into account a 3N force. There are
two other important motivations behind this work. The first
one is to show in details that the  HH expansion applied to the
four--nucleon bound and scattering problems is
very powerful even for  realistic NN interactions. The second
motivation is the possibility of the
extension of the method to larger systems. The feasibility of such an
application would require the solution of the following different
problems. First, the calculation of the generalized ``Raynal-Revai''
coefficients, namely, of the coefficients relating HH functions
constructed with different sets of Jacobi vectors. The 
direct generalization of the algorithm proposed in Ref.~\cite{V98} 
is adequate for $A=5\div 8$. Otherwise, different algorithms could be
used~\cite{NK94,DLWD95,E95}. Second, the computation of the matrix 
elements of NN and 3N interactions, which can be reduced to the
evaluation of low-dimensional integrals as in the present case.
In particular, the possibility of approximating the 3N interaction 
as acting only on HH functions of low $K$, as discussed in 
Sec.~\ref{sec:trun}, should appreciably simplify this task.
Finally, the choice of an optimal subset of
HH functions. As $A$ grows, the number of HH states for a given $K$
increases very rapidly. The criteria for selecting the subclasses of HH
functions chosen  in the present paper can be readily generalized to
systems with $A>4$. However, additional properties of the HH function
could be exploited to further reduce the number of terms in the expansion.
For example, one could take into account the symmetry of the space
part of the states constructed as a product of the HH functions 
and the spin-isospin  states. Another possibility to be explored is to
include classes of  HH functions constructed with those Jacobi vectors
pertaining to different partitions of the particles. For example, in
the study of the $d+\tri\rightarrow \heq+p$ reaction, the use of HH
functions constructed in terms of $2+3$ and $4+1$ clusterizations
should be very useful.

\section{Acknowledgments.}
The author would like to thank Prof. R. Schiavilla and
Prof. L. Lovitch for useful discussions and a critical 
reading of the manuscript.

\appendix
\section*{Appendix}

In this Appendix, the method used for
calculating the matrix elements of a local 3N interaction operator $H_{3N}$
between the antisymmetric hyperangular--spin--isospin states defined 
in Eq.~(\ref{eq:PSI}) will be briefly illustrated. The major problem 
to be overcome is to achieve a sufficient numerical precision, so that 
the differential equations for the functions $u_{KLST,\mu}(\rho)$
defined in Eq.~(\ref{eq:PSI3}) could be solved without any numerical 
trouble. In general, a 3N interaction is written as follows
\begin{equation}
   H_{3N}=\sum_{i<j<k}\; \sum_{cyclic} \; W_{3N}(i;j,k)\ ,
\end{equation}
where $\sum_{cyclic}$ represents a cyclic sum over induces $i$, 
$j$ and $k$ and $W_{3N}(i;j,k)$ is symmetric under the exchange of
the particles $j$ and $k$. Therefore, the problem can be reduced to the 
computation of the matrix 
element of the operator $W_{3N}(1;2,3)$.
Once the antisymmetric hyperangular--spin--isospin states 
are expanded in terms of the jj states as in 
Eq.~(\ref{eq:PSI3jj}), one has to compute the following integrals:
\begin{equation}
 {\cal W}_{\nu,\nu'}(\rho)=\int d\Omega\; 
  \langle\Xi^{KTJ\pi}_\nu(1,2,3,4)| \;
  W_{3N}(1;2,3) \; 
  |\Xi^{K^\prime TJ\pi}_{\nu^\prime}(1,2,3,4)\rangle\ ,
  \label{eq:3N1}
\end{equation}
where the states $\Xi^{KTJ\pi}_\nu(1,2,3,4)$ are defined in
Eq.~(\ref{eq:PHIjj}) (hereafter all the Jacobi vectors are 
chosen to correspond to the permutation $p=1$, and therefore 
the index $p$ will be omitted).
In the case of the Urbana or TM-like 3N interactions,
$W_{3N}(1;2,3)$ can be taken to have the general form~\cite{CPW83}
\begin{equation}
  W_{3N}(1;2,3)=\sum_{p,q=1}^6 F_{p,q}(r_{12},r_{13},\mu_{12,13}) 
  \; {\cal O}^p_{12} \; {\cal O}^q_{13} \ ,  \label{eq:3N2a}
\end{equation}
where $\mu_{12,13}=
  \widehat \bfr_{12}\cdot\widehat \bfr_{13}$ and $\bfr_{ij}$ 
is the relative distance between the particles $i$ and $j$.
In the latter equation, ${\cal O}^{p=1,6}_{ij}$ are the operators
\begin{equation}
  {\cal O}^{p=1,6}_{ij}=1,\; (\bmta_i\cdot\bmta_j), \;
   (\bmsi_i\cdot\bmsi_j),\;  
   (\bmsi_i\cdot\bmsi_j) (\bmta_i\cdot\bmta_j), \;
   r_{ij}^2S_{ij}, \; 
    r_{ij}^2S_{ ij} (\bmta_i\cdot\bmta_j)\ ,\label{eq:3N2b}
\end{equation}
where $S_{ij}$ is the tensor operator (the factor $r_{ij}^2$ has been 
included in the definition of ${\cal O}^{p=5,6}$ so that these operators
are polynomials in the Cartesian coordinates of the particles).

Since $W_{3N}(1;2,3)$ depends only on the variables
$\rho,\hypfi_3,\hypfi_2,\widehat\jac_2,\widehat\jac_3$ we can easily
integrate over the variables $\widehat\jac_1$. Moreover, by evaluating the
spin-isospin traces and the integrals over the angles
$\widehat\jac_2$, $\widehat\jac_3$ (except for
$\mu=\widehat\jac_2\cdot\widehat\jac_3$) one reduces the matrix element given
in Eq.~(\ref{eq:3N1}), to an integral of the type:
\begin{equation}
 {\cal W}_{\nu,\nu'}(\rho)= \sum_{p,q} \int_{-1}^1 dz\; \sqrt{1+z}
    \int_{-1}^1 dx\; \sqrt{1-x^2} \int_{-1}^1 d\mu\; 
    F_{p,q}(r_{12},r_{13},\mu_{12,13})  {\cal P}_{p,q}(z,x,\mu)
    \, \label{eq:3N3}
\end{equation}
where
\begin{equation}
  z=\cos 2\hypfi_{3}=2 {r_{12}^2\over \rho^2}-1\ , \qquad 
  x=\cos 2\hypfi_{2}=2 {\jac_2^2\over\rho^2}-1 \ , 
  \label{eq:3N4}
\end{equation}
and
\begin{equation}
   {\cal P}_{p,q}(z,x,\mu) = {(4\pi)^2\over 128\sqrt{2}}
    \int d\widehat\jac_1\;
   \langle \Xi^{KTJ\pi}_\nu(1,2,3,4) |
   {\cal O}^p_{12} \; {\cal O}^q_{13} 
   | \Xi^{K^\prime TJ\pi}_{\nu^\prime}(1,2,3,4)\rangle\ .
   \label{eq:3N5}
\end{equation}
In Eq.~(\ref{eq:3N5}) the integration over the angles 
$\widehat\jac_2$, $\widehat\jac_3$ (except for
$\mu=\widehat\jac_2\cdot\widehat\jac_3$) and the trace over the
spin-isospin degrees of freedom is implicit. This latter part of
the calculation can be performed analytically in terms of 
Wigner $D$-matrices and Clebsh-Gordan coefficients.
The remaining  two-dimensional integration over $d\widehat\jac_1=
d\cos\theta_1\;d\phi_1$ in Eq.~(\ref{eq:3N5}) 
can be easily performed by taking into account that the
integrand is a polynomial in $\cos\theta_1$ and $\cos\phi_1$
of degree $K+K^\prime$.

The functions ${\cal P}$ can be therefore calculated {\it exactly} using 
an appropriate Gauss integration formula with a small number of points.
On the other hand, the functions $F_{p,q}$ entering the 
tri-dimensional integral~(\ref{eq:3N3}) are very complicated 
functions of the variables $z,x,\mu$.
Therefore, the integral~(\ref{eq:3N3}) requires the use of extended 
and dense integration grids (about $(1000)^3$ points) so as to yield
the needed accuracy. Since, the same integration has to be repeated
for each $\nu$, $\nu'$, the complete calculation
of ${\cal W}_{\nu,\nu'}$ could be very time consuming.

However, the function ${\cal P}_{p,q}(z,x,\mu)$ 
can be written in general as
\begin{equation}
  {\cal P}_{p,q}(z,x,\mu)={\cal P}^e_{p,q}(z,x,\mu)+\sqrt{1+x}\;
  \sqrt{1-z^2}\;
   {\cal P}^o_{p,q}(z,x,\mu)\ ,\label{eq:3N7}
\end{equation}
where ${\cal P}^e_{p,q}$ and ${\cal P}^o_{p,q}$ are 
polynomials in $z$, $x$ and $\mu$
of maximum degree $N=K+K^\prime+2$, the $2$ coming (eventually) from 
the factor $r_{ij}^2$ multiplying the tensor operators 
in Eq.~(\ref{eq:3N2b}).
More precisely ${\cal P}^e$ ($\sqrt{1+x}\sqrt{1-z^2}{\cal P}^o$) 
is the even (odd) part 
of ${\cal P}$ with respect to the variable $\mu$.

Now, if $p(t)$ is a polynomial of degree $n$ with respect to the
variable $t$ and 
its value in each of $n+1$ points $t_1,\ldots,t_{n+1}$ is known,
using the following  ``Lagrange interpolation'' formula, $p(t)$ 
can be computed {\it exactly}  for any $t$:
\begin{equation}
  p(t)=\sum_{i=1,\ldots,n+1} p(t_i) L_i^{(n+1)}(t)\ ,
   \qquad L^{(n+1)}_i(t)\equiv \prod_{j=1,\ldots,n+1}^{j\ne i}
    {t-t_j\over t_i-t_j}\ .\label{eq:3N6}
\end{equation}
Therefore, once three sets of points $z_1,\ldots,z_{N+1}$,
$x_1,\ldots,x_{N+1}$ and $\mu_1,\ldots,\mu_{N+1}$ in the interval $(-1,1)$
have been selected,
and ${\cal P}^{e,o}_{p,q}$ in the $(N+1)^3$ points $z_i,x_j,\mu_k$
have been computed,
the functions ${\cal P}_{p,q}$ are then known {\it exactly} for all 
possible values of $(z,x,\mu)$. Since the matrix elements of the 
3N interaction are needed only between HH states with $K\lesssim 30$
(and therefore $\max[N]\ll 100$), as discussed in Sec.~\ref{sec:trun}, 
this means that in practice the functions ${\cal P}$ have to be evaluated 
only a fairly small number of times. Finally, if we evaluate
\begin{eqnarray}
  I^{p,q}_{i,j,k}&=& \int_{-1}^1 dz\; \sqrt{1+z}
    \int_{-1}^1 dx\; \sqrt{1-x^2} \int_{-1}^1 d\mu\; 
    F_{p,q}(r_{12},r_{13},\mu_{12,13}) \times\nonumber \\
    && \qquad \qquad \times 
    L^{(N+1)}_i(z) L^{(N+1)}_j(x) L^{(N+1)}_k(\mu)
    \ , \label{eq:3N9}\\
\noalign{\medskip}
  J^{p,q}_{i,j,k}&=& \int_{-1}^1 dz\; \sqrt{1+z}
    \int_{-1}^1 dx\; \sqrt{1-x^2} \int_{-1}^1 d\mu\; 
    F_{p,q}(r_{12},r_{13},\mu_{12,13})\times\nonumber \\
  && \qquad\qquad \times \sqrt{1+x}\;\sqrt{1-z^2} \;
    L^{(N+1)}_i(z) L^{(N+1)}_j(x) L^{(N+1)}_k(\mu)
    \ , \label{eq:3N10}
\end{eqnarray}
the required matrix elements can be obtained simply as
\begin{equation}
     {\cal W}_{\nu,\nu'}(\rho)=\sum_{p,q} \sum_{i,j,k=1}^{N}  
     \Bigl[{\cal P}^e_{p,q}(z_i,x_j,\mu_k)
     I^{p,q}_{i,j,k} +  {\cal P}^o_{p,q}(z_i,x_j,\mu_k)
     J^{p,q}_{i,j,k} \Bigr ]   \ . \label{eq:3N8}
\end{equation}
where
the integrals given in Eqs.~(\ref{eq:3N9}) and~(\ref{eq:3N10})
do not depend any more on the quantum numbers $\nu$, $\nu'$ of the 
HH states  and
therefore can be computed with the necessary accuracy
once and for all and stored on computer disks. 
In this way, the matrix elements ${\cal W}_{\nu,\nu'}(\rho)$ 
obtained via Eq.~(\ref{eq:3N8}) are obtained very quickly (with 
only $\sim (N+1)^3$ operations).

\newpage

\begin{table}
\caption[Table]{\label{tb:nhh}
Number of four-nucleon antisymmetrical hyperspherical--spin--isospin states 
for the case $J=0$, $T=0$ and $\pi=+$ and the selected values of
the grandangular quantum number $K$ and total angular momentum $L$.
$M_{KL}$ is the total number of the states defined in 
Eq.~(\ref{eq:PSI}).  $M^\prime_{KL}$ gives the number of the
linearly independent states with $\ell_1+\ell_2+\ell_3\leq 6$. See the
text for details.
}
\begin{tabular}{r@{$\qquad$}r@{$\qquad$}r@{$\qquad$}r@{$\qquad$}
                r@{$\qquad$}r@{$\qquad$}r}
    $K$   & \multicolumn{2}{c}{$L=0$}  &
            \multicolumn{2}{c}{$L=1$}  &
            \multicolumn{2}{c}{$L=2$}  \\
    & $M_{KL}$ & $M_{KL}^\prime $
    & $M_{KL}$ & $M_{KL}^\prime $
    & $M_{KL}$ & $M_{KL}^\prime $  \\
\hline
    0  &    2 &   1 &       &     &       &     \\
    2  &   10 &   1 &    9  &   1 &    6  &  1  \\
    4  &   30 &   4 &   45  &   4 &   30  &   3 \\
    6  &   70 &   8 &  135  &  12 &   89  &   9 \\
    8  &  140 &  14 &  315  &  27 &  205  &  18 \\
   10  &  252 &  24 &  630  &  54 &  405  &  36 \\
   12  &  420 &  41 & 1,134 &  96 &  721  &  63 \\
   14  &  660 &  59 & 1,890 & 160 & 1,190 & 102 \\
   16  &  990 &  90 & 2,970 & 250 & 1,854 & 158 \\
   18  & 1,430& 128 & 4,455 & 375 & 2,760 & 236 \\
   20  & 2,002& 176 & 6,435 & 488 & 3,960 & 321 \\
   22  & 2,730& 235 & 9,009 & 585 & 5,511 & 385 \\
   24  & 3,640& 282 &12,285 & 675 & 7,475 & 445 \\
   30  &  3,876  &  & 9,180  &  & 16,540 &   \\
   40  &  10,626 &  & 26,565 &  & 47,145 &   \\
   50  &  23,751 &  & 61,425 &  & 107,900&
\end{tabular}
\end{table}

\newpage

\begin{table}
\caption[Table]{\label{tb:chan0p0}
Quantum numbers of the first channels considered in the expansion of
the $\alpha$-particle WF. See the text for details.
}
\begin{tabular}{r@{$\qquad$}r@{$\qquad$}r@{$\qquad$}r@{$\qquad$}r@{$\qquad$}
                r@{$\qquad$}r@{$\qquad$}r@{$\qquad$}r@{$\qquad$}r@{$\qquad$}
                r@{$\qquad$}r}
$\alpha$ & $\ell_1$ & $\ell_2$ & $\ell_3$ & $L_2$ & $L$ &
$S_a$ & $S_b$ & $S$ & $T_a$ & $T_b$ & $T$ \\
\hline
 1&0&0&0&0&0& 1&1/2& 0& 0&1/2&0 \\
 2&0&0&0&0&0& 0&1/2& 0& 1&1/2&0 \\
 3&0&0&2&0&2& 1&3/2& 2& 0&1/2&0 \\
 4&1&1&0&0&0& 1&1/2& 0& 0&1/2&0 \\
 5&1&1&0&0&0& 0&1/2& 0& 1&1/2&0 \\
 6&1&1&0&1&1& 1&1/2& 1& 0&1/2&0 \\
 7&1&1&0&1&1& 1&3/2& 1& 0&1/2&0 \\
 8&1&1&0&1&1& 0&1/2& 1& 1&1/2&0 \\
 9&0&2&0&2&2& 1&3/2& 2& 0&1/2&0 \\
10&2&0&0&2&2& 1&3/2& 2& 0&1/2&0 \\
11&1&1&0&2&2& 1&3/2& 2& 0&1/2&0 \\
12&1&0&1&1&0& 1&1/2& 0& 1&1/2&0 \\
13&1&0&1&1&0& 0&1/2& 0& 0&1/2&0 \\
14&0&1&1&1&0& 1&1/2& 0& 1&1/2&0 \\
15&0&1&1&1&0& 0&1/2& 0& 0&1/2&0 \\
16&1&0&1&1&1& 1&1/2& 1& 1&1/2&0 \\
17&1&0&1&1&1& 1&3/2& 1& 1&1/2&0 \\
18&1&0&1&1&1& 0&1/2& 1& 0&1/2&0 \\
19&0&1&1&1&1& 1&1/2& 1& 1&1/2&0 \\
20&0&1&1&1&1& 1&3/2& 1& 1&1/2&0 \\
21&0&1&1&1&1& 0&1/2& 1& 0&1/2&0 \\
22&1&0&1&1&2& 1&3/2& 2& 1&1/2&0 \\
23&0&1&1&1&2& 1&3/2& 2& 1&1/2&0 \\
\hline
\end{tabular}
\end{table}

\newpage

\begin{table}
\caption[Table]{\label{tb:conv}
Convergence of $\alpha$--particle binding energies (MeV) corresponding
to the inclusion in the WF of the different classes C1--C6
in which the HH basis has been subdivided.}
\begin{tabular}{c@{$\quad$}c@{$\quad$}c@{$\quad$}c@{$\quad$}
                c@{$\quad$}c@{$\quad$}|@{$\quad$}
                c@{$\qquad$}c@{$\qquad$}c@{$\ $}}
$K_1$ & $K_2$ & $K_3$ & $K_4$ & $K_5$ & $K_6$ &
  MT-V   & AV18 & AV18+UIX \\
\hline
 20 &&&&&& 28.928  & 14.701  & 14.902      \\
 30 &&&&&& 29.794  & 15.992  & 16.162     \\
 40 &&&&&& 29.962  & 16.172  & 16.337     \\
 50 &&&&&& 30.008  & 16.205  & 16.369     \\
 60 &&&&&& 30.024  & 16.213 & 16.377     \\
 70 &&&&&& 30.032  & 16.214 & 16.379     \\
 72 &&&&&& 30.033  & 16.214 & 16.379     \\
\hline
 72 & 8  &&&&& 30.714  & 18.286  & 18.985      \\
 72 & 16 &&&&& 31.170  & 19.755  & 20.645      \\
 72 & 24 &&&&& 31.240  & 19.967  & 20.865      \\
 72 & 32 &&&&& 31.256  & 20.014  & 20.909     \\
 72 & 36 &&&&& 31.259  & 20.022 & 20.916     \\
 72 & 40 &&&&& 31.261  & 20.026 & 20.919     \\
\hline
 72 & 40 &  8  &&&& 31.300 & 21.940 & 24.682   \\
 72 & 40 &  16 &&&& 31.336 & 23.237 & 27.142  \\
 72 & 40 &  24 &&&& 31.340 & 23.371 & 27.350  \\
 72 & 40 &  30 &&&& 31.341 & 23.385 & 27.370 \\
 72 & 40 &  34 &&&& 31.341 & 23.388 & 27.373 \\
\hline
 72 & 40 &  34 &  8  &&& 31.341 & 23.525 & 27.553 \\
 72 & 40 &  34 &  16 &&& 31.344 & 24.086 & 28.312 \\
 72 & 40 &  34 &  20 &&& 31.346 & 24.145 & 28.382 \\
 72 & 40 &  34 &  24 &&& 31.347 & 24.163 & 28.404 \\
 72 & 40 &  34 &  28 &&& 31.347 & 24.170 & 28.414 \\
\hline
 72 & 40 &  34 &  28 & 16 &&    & 24.181 &  28.427 \\
 72 & 40 &  34 &  28 & 20 &&    & 24.191 &  28.439 \\
 72 & 40 &  34 &  28 & 24 &&    & 24.195 &  28.444 \\
\hline
 72 & 40 &  34 &  28 & 24 & 4  && 24.205 &  28.456 \\
 72 & 40 &  34 &  28 & 24 & 8  && 24.209 &  28.461 \\
 72 & 40 &  34 &  28 & 24 & 12 && 24.210 &  28.462  \\
\hline
\hline
\multicolumn{6}{c}{``exact''}   & 31.360  & 24.25  & 28.50 \\
\hline
\hline
\end{tabular}
\end{table}

\newpage

\begin{table}
\caption[Table]{\label{tb:estra}
Increments of the $\a$-particle BE $\bar\Delta(K)$, computed using
Eqs.~(\ref{eq:c2diffb}) for the various classes $i=1,\ldots,6$ 
and the MT-V and AV18 potential models. The quantities $c(K,p)$ 
are defined in
Eq.~(\ref{eq:extra}) and $(\Delta B)_i$, given by $c(k,p)
\bar\Delta_i(K)$, represents the ``missing BE'' for having 
truncated the expansion over the class $i$ up to
the given value of $K=K_M$. Finally, the ``total missing BE'' 
$(\Delta B)_T$ is computed from Eq.~(\ref{eq:dbet}).
}
\begin{tabular}{l @{$\qquad$} c @{$\qquad$} c @{$\qquad$} c
    @{$\qquad$} c @{$\qquad$} c @{$\qquad$} c @{$\qquad$} c}
\hline
 & & \multicolumn{3}{c}{MT-V} &  \multicolumn{3}{c}{AV18}    \\
\hline
 $i$   & $K_M$  &  $\bar\Delta_i(K)$ [keV] & $c(K,5)$ & $(\Delta B)_i$
 [keV] &  $\bar\Delta_i(K)$ [keV] & $c(K,7)$ & $(\Delta B)_i$ [keV]\\
\hline
 1 & 72      & 0.89 & 8.51 & 7.57  & 0.10 & 5.52 & 0.55\\
 2 & 40      & 0.71 & 4.52 & 3.21  & 0.91 & 2.86 & 2.60 \\
 3 & 34      & 0.07 & 3.77 & 0.26  & 1.16 & 2.37 & 2.75 \\
 4 & 28      & 0.13 & 3.03 & 0.39  & 2.30 & 1.87 & 4.30 \\
 5 & 24      &  -   & -    & -     & 1.47 & 1.54 & 2.26 \\
\hline
\multicolumn{4}{l}{$(\Delta B)_T$}   & 11.43 & & & 12.46\\
\hline
\end{tabular}
\end{table}

\newpage
\begin{table}
\caption[Table]{\label{tb:res}
The $\alpha$--particle binding energies $B$ (MeV),
rms radii (fm) and expectation value $\langle K\rangle$ 
of the kinetic energy operator
(MeV) for various central interaction models as computed by 
means of the HH expansion are compared with the results obtained 
by other techniques. The binding energies obtained by
using the extrapolation technique described
in Sect.~\ref{sec:conv} are enclosed in parentheses.
} 
\begin{tabular}{l@{$\qquad$}l@{$\qquad$}c@{$\quad$}c@{$\quad$}c}
Interaction   & Method & $B$ & $\langle r^2\rangle^{1/2}$ & 
 $\langle K\rangle$  \\
\hline
Volkov
    & HH~(this work)          & 30.420\pp & 1.490  & 50.319  \\
    & SVM~\protect\cite{VS95} & 30.42\ppp  & 1.49   &  \\
    & HH~\protect\cite{B81}   & 30.399\pp &        &  \\
\hline
ATS3
    & HH~(this work)           & 31.618\pp   & 1.412 & 74.366 \\
    & SVM~\protect\cite{V04}   & 31.616\pp   & 1.42   &        \\
\hline
Minnesota
    & HH~(this work)           & 29.947\pp & 1.4105 & 58.086 \\
    & SVM~\protect\cite{VS95}  & 29.937\pp & 1.41   &        \\
    & EIHH~\protect\cite{BLO00}& 29.96\ppp & 1.4106 &        \\
\hline
MT-V
    & HH~(this work)           & 31.347(31.358) & 1.4081  & 69.792  \\
    & SVM~\protect\cite{VS95}  & 31.360\pp & 1.4087  &       \\
    & EIHH~\protect\cite{BLO00}& 31.358\pp & 1.40851 &       \\
    & CRCG~\protect\cite{KK90} & 31.357\pp &         &       \\
    & FY~\protect\cite{KG92}   & 31.36\ppp  &         &       \\
    & ATMS~\protect\cite{A86}  & 31.36\ppp  & 1.40    &       \\
\hline
MT-I/III
    & HH~(this work)          & 30.310(30.331)   & 1.4380 & 66.180  \\
    & FY~\protect\cite{SSK92} & 30.312\pp  &        &  \\
\hline
\end{tabular}
\end{table}

\newpage

\begin{table}
\caption[Table]{\label{tb:resB}
The $\alpha$--particle binding energies $B$ (MeV), the rms radii (fm), 
the expectation values of the kinetic energy operator $\langle K\rangle$
(MeV), and the $P$ and $D$ probabilities (\%) for various realistic 
interaction models as computed by means of the HH expansion
are compared with the results obtained by other techniques.
The binding energies obtained by
using the extrapolation technique described
in Sect.~\ref{sec:conv} are enclosed in parentheses.}
\begin{tabular}{l@{$\quad$}l@{$\qquad$}c@{$\quad$}
                c@{$\quad$}c@{$\quad$}c@{$\quad$}c}
Interaction   & Method & $B$ & $\langle r^2\rangle^{1/2}$ &
              $\langle K\rangle$  & $P_P$ & $P_D$ \\
\hline
AV18
    & HH~(this work)           & 24.210(24.222) 
      &  97.84 & 1.512  & 0.347  & 13.74 \\
    & FY~\protect\cite{Nea02}  & 24.25\ppp  
      &  97.80 &        & 0.35   & 13.78 \\
    & FY~\protect\cite{LC04}   & 24.223\pp 
       & 97.77 & 1.516  &    &  \\
\hline
Nijm~II
    & HH~(this work)     & 24.419(24.432)  
      & 100.27  & 1.504 & 0.334 & 13.37 \\
    & FY~\protect\cite{Nea02}  & 24.56\ppp    & 100.31  & &  &  \\
\hline
AV18+UIX
    & HH~(this work)           & 28.462(28.474) 
      & 113.30 & 1.428 & 0.73  & 16.03  \\
    & FY~\protect\cite{Nea02}  & 28.50\pp   
      & 113.21 & & 0.75  & 16.03 \\
    & GFMC~\protect\cite{Wea00}& 28.34(4)\phantom{28.47}
      & 110.7(7) & 1.44 & & \\
\hline
AV18+TM'
    & HH~(this work)     & 28.301(28.313) 
      & 110.27 & 1.435 & 0.73  & 15.63  \\
    & FY~\protect\cite{Nea02}  & 28.36\ppp
         & 110.14 & & 0.75  & 15.67 \\
\hline
\end{tabular}
\end{table}

\newpage

\begin{table}
\caption[Table]{\label{tb:isom}
Percentages of the total isospin components $T=1$ and $2$ in
the $\alpha$-particle ground states for various
interaction models.}
\begin{tabular}{l@{$\quad$}l@{$\;$}|@{$\;$}c@{$\quad$}c}
 interaction  & method & $P_{T=1}$ [\%] &  $P_{T=2}$ [\%] \\
\hline
 AV18     & HH~{this work}          & $2.8\; 10^{-3}$ & $5.2\; 10^{-3}$ \\
 AV18     & FY~\protect\cite{Nea02} & $3\;   10^{-3}$ & $5\; 10^{-3}$ \\
 Nijm-II  & HH~{this work}          & $1.6\; 10^{-3}$ & $7.4\; 10^{-3}$ \\
 AV18+UIX & HH~{this work}          & $2.5\; 10^{-4}$ & $5.0\; 10^{-3}$ \\
\hline
\end{tabular}
\end{table}

\begin{table}
\caption[Table]{\label{tb:isom2}
Effect of the inclusion of the various isospin mixing terms in the
nuclear Hamiltonian on the percentages of the total isospin components
$T=1$ and $2$ in the $\alpha$-particle ground states. The calculations
have been performed using the AV18 model for the nuclear Hamiltonian. For 
the explanation of the various terms $H_{IC}$, etc, see the text.
}
\begin{tabular}{l@{$\;$}|@{$\;$}c@{$\quad$}c}
 interaction   & $P_{T=1}$ [\%] &  $P_{T=2}$ [\%] \\
\hline
 $H_{IC}$                    & 0 & 0 \\
 $H_{IC}+H_C$                & $1.5\; 10^{-3}$ & $0.1\; 10^{-3}$ \\
 $H_{IC}+H_C+H_{CSB}$        & $3.0\; 10^{-3}$ & $4.9\; 10^{-3}$ \\
 $H_{IC}+H_C+H_{CSB}+H_{em}$ & $2.8\; 10^{-3}$ & $5.2\; 10^{-3}$ \\
\hline
\end{tabular}
\end{table}

\newpage

\begin{table}
\caption[Table]{\label{tb:truj}
Effects of the truncation of the NN potential when acting only on 
pairs having total angular momentum $j\le j_{\rm M}$.
The potential model chosen is the AV18 and the
selected HH basis has $\{
K_1,K_2,K_3,K_4,K_5,K_6\}=\{64,40,34,24,0,0\}$ .}
\begin{tabular}{c@{$\;$}|@{$\;$}c@{$\quad$}c@{$\quad$}c@{$\quad$}c}
 $j_{\rm M}$   & $B$ (MeV) & $T$ (MeV) & $P_P$ (\%) &  $P_D$ (\%)  \\
\hline
 4      & -24.124 & 97.692 & 0.344 & 13.713 \\
 6      & -24.161 & 97.771 & 0.345 & 13.724 \\
 8      & -24.164 & 97.773 & 0.345 & 13.725 \\
10      & -24.165 & 97.773 & 0.345 & 13.725 \\
20      & -24.163 & 97.774 & 0.345 & 13.720 \\
$\infty$& -24.163 & 97.774 & 0.345 & 13.720 \\
\hline
\end{tabular}
\end{table}

\begin{table}
\caption[Table]{\label{tb:truk}
Effects of the truncation of the 3N potential when acting only on
four-body HH  states
of grand angular quantum number  $K\le K_{\rm M}$.
The potential model chosen is the AV18+UIX and the
selected HH basis has $\{
K_1,K_2,K_3,K_4,K_5,K_6\}=\{64,40,34,0,0,0\}$ .}
\begin{tabular}{c@{$\;$}|@{$\;$}c@{$\quad$}c@{$\quad$}c@{$\quad$}c}
 $K_{\rm M}$   & $B$ (MeV) & $T$ (MeV) & $P_P$ (\%) &  $P_D$ (\%)  \\
\hline
20 & 27.351 & 109.71 & 0.596 & 15.05 \\
24 & 27.366 & 109.67 & 0.596 & 15.05 \\
30 & 27.372 & 109.65 & 0.596 & 15.05 \\
34 & 27.374 & 109.64 & 0.596 & 15.05 \\
\hline
\end{tabular}
\end{table}

\newpage

\begin{table}
\caption[Table]{\label{tb:anc}
ANC's and the parameter $D_2^{dd}$ obtained with the HH expansion 
and the solution of the differential equation of Eq.~(\ref{eq:anc7}) 
for two potential models. The $D_2$ parameter is 
defined in Eq.~(\ref{eq:anc9}). In the third row, the theoretical
estimate for $D_2^{dd}$ of Ref.~\cite{adhi94} is also shown. Finally, 
in the last three rows, some available experimental value for the
parameter $D_2^{dd}$ have been also reported.
}
\begin{tabular}{l@{$\quad$}c@{$\quad$}c@{$\quad$}c@{$\quad$}
                c@{$\quad$}c@{$\quad$}}
Parameter & $C_S^{pt}$ & $C_S^{nh}$ & $C_S^{dd}$ & $C_D^{dd}$ & 
$D_2^{dd}$ (fm$^2$) \\
\hline
AV18     &  1.72 & 1.67 & 1.96  & $-0.209$ & $-0.115$ \\
AV18+UIX &  1.75 & 1.69 &  1.99 & $-0.277$ & $-0.113$ \\
Adhikari {\it et al.} \protect\cite{adhi94}
       & & & & & $-0.12$ \\
\hline
Karp {\it et al.} \protect\cite{karp86}
       & & & & & $-0.3\pm 0.1$ \\
Merz {\it et al.} \protect\cite{merz87}
       & & & & & $-0.19 \pm 0.04$ \\
Weller {\it et al.} \protect\cite{weller86}
       & & & & & $-0.2 \pm 0.05$ \\
\hline
\end{tabular}
\end{table}

\newpage

\begin{thebibliography}{99}
\bibitem{V04} K. Varga, private communications (please note that
the SVM result reported in Ref.~\cite{VS95} 
in Table~V has been probably computed with a different version of 
the ATS3 potential).
\end{thebibliography}
\end{document}